\documentclass[aps,prl,twocolumn,showkeys,noprintnumbers,amssymb,aps,superscriptaddress,longbibliography]{revtex4-2}
\usepackage{graphicx}
\usepackage{dcolumn}
\usepackage{bm}
\usepackage{color}
\usepackage{epstopdf}
\usepackage{epsfig}
\usepackage{xcolor}
\usepackage[T1]{fontenc}
\usepackage{amsmath}
\usepackage[breaklinks,colorlinks,bookmarks=false,citecolor=blue,linkcolor=red,urlcolor=blue]{hyperref}
\usepackage{epstopdf}
\DeclareUnicodeCharacter{200B}{}
\begin{document}


\title{Approaching Kasteleyn transition in frustrated quantum Heisenberg antiferromagnets}
\author{Katar\'ina Kar{l}'ov\'a}
\email{katarina.karlova@upjs.sk}
\affiliation{Laboratoire de Physique Th\'eorique et
Mod\'elisation, CNRS UMR 8089, CY Cergy Paris Universit\'e, Cergy-Pontoise, France}
\affiliation{Institute of Physics, Faculty of Science, P. J. \v{S}af\'{a}rik University, Park Angelinum 9, 04001 Ko\v{s}ice, Slovakia}
\author{Afonso Rufino}
\affiliation{Institute of Physics, Ecole Polytechnique F\'ed\'erale de Lausanne (EPFL), CH-1015 Lausanne, Switzerland}
\author{Taras Verkholyak}
\affiliation{Yukhnovskii Institute for Condensed Matter Physics of the National Academy of Sciences of Ukraine,
	Svientsitskii Street 1, 79011 L’viv, Ukraine}
\affiliation{Professor Ivan Vakarchuk Department for Theoretical Physics, Ivan Franko National University of Lviv, 12 Drahomanov Street, Lviv, Ukraine}
\author{Nils Caci}
\affiliation{Laboratoire Kastler Brossel, \'Ecole Normale Sup\'erieure – Universit\'e PSL, CNRS, Sorbonne Universit\'e, Coll\'ege de France, 75005 Paris, France}
\author{Stefan Wessel}
\affiliation{Institute for Theoretical Solid State Physics, JARA FIT, and JARA CSD, RWTH Aachen
University, 52056 Aachen, Germany}
\author{Jozef Stre\v{c}ka}
\affiliation{Institute of Physics, Faculty of Science, P. J. \v{S}af\'{a}rik University, Park Angelinum 9, 04001 Ko\v{s}ice, Slovakia}
\author{Fr\'ed\'eric Mila}
\affiliation{Institute of Physics, Ecole Polytechnique F\'ed\'erale de Lausanne (EPFL), CH-1015 Lausanne, Switzerland}
\author{Andreas Honecker}
\affiliation{Laboratoire de Physique Th\'eorique et
Mod\'elisation, CNRS UMR 8089, CY Cergy Paris Universit\'e, Cergy-Pontoise, France}

\date{\today}

\begin{abstract}
We show that the Kasteleyn transition, the abrupt proliferation of infinite strings of defects in classical dimer and related models, can also be relevant for frustrated 2d quantum magnets. This is explicitly demonstrated in a phase of the spin-1/2 Heisenberg diamond-decorated honeycomb lattice where a family of exact eigenstates built as products of dimer and plaquette singlets can be mapped onto the dimer coverings of the honeycomb lattice. The low-temperature properties of this phase are accurately described by an effective dimer model with anisotropic activities and a small, tunable density of monomers, leading to an arbitrarily sharp crossover version of the Kasteleyn transition. The generalization to other geometries and the possibility to realize this model in organo-metallic compounds are briefly discussed.\end{abstract}
\keywords{Heisenberg antiferromagnet, spin frustration, dimer model, Kasteleyn transition, string excitations, exact diagonalization, corner transfer matrix renormalization group, quantum Monte Carlo}

\maketitle

\paragraph{Introduction}
Dimer models play a central role in statistical mechanics, combinatorics, and frustrated magnetism, offering paradigms for exotic phases and critical phenomena in two dimensions \cite{Kasteleyn1961,TemperleyFisher1961,nagle1989,Wu2006, kenyon2009,Henley2010,Moessner2011,shah2025}. On bipartite lattices, close-packed dimer coverings exhibit algebraic correlations and topological constraints that mimic emergent gauge fields and Coulomb phases \cite{Henley2010}. One of the most intriguing aspects of hard-dimer models is the Kasteleyn transition, first identified in a classical dimer model on the honeycomb lattice \cite{Kasteleyn1963}, where anisotropic bond weights induce a singularity in the specific heat without conventional symmetry breaking \cite{Wu1968, Nagle1975}.

Over the years, dimer models have proven useful far beyond their combinatorial origin. Exact mappings relate them to vertex models \cite{Lieb1967sixvertex}, Ising systems \cite{Fisher1966,Lieb1967}, and ferroelectric-type materials \cite{Wu1967, Wu1968}. More recently, dimer-based descriptions have been extended to constrained quantum magnets, where valence-bond coverings, emergent monomer defects, and ground-state degeneracy lead to rich thermodynamic and topological behavior \cite{shah2025,Misguich2002,Morita2016,Mahapatra2024,Powell2,Powell3}, and the emergence of infinite string excitations in quantum dimer models has been identified as a hallmark of Kasteleyn criticality~\cite{Powell2022}. However, traces of Kasteleyn physics have not been identified so far in SU(2) Heisenberg quantum magnets, while they have been observed in strongly anisotropic, Ising-like magnetic systems such as spin-ice materials under specific orientations of magnetic field \cite{spinice1, spinice2, spinice3}. More generally, thermal phase transitions associated to SU(2) symmetry breaking are strictly forbidden in 2d by the Mermin–Wagner theorem \cite{MerminWagner}, highlighting the challenge of realizing unconventional finite-temperature criticality in such systems. 

Here, we close this gap by demonstrating how the physics of the Kasteleyn transition can be realized in Heisenberg quantum magnets: we examine a spin-$\tfrac{1}{2}$ Heisenberg model on a diamond-decorated honeycomb lattice and show that it realizes an emergent dimer phase with thermodynamic signatures characteristic of the Kasteleyn transition. Heisenberg models on diamond-decorated planar lattices have attracted considerable recent attention due to the emergence of exotic quantum phases as well as quantum and thermal phase transitions \cite{Morita2016,Hirose2016,Hirose2017b,Hirose2018,Hirose2020,Caci2023,Karlova2024,DmitrievPRB,DmitrievZNA,KS-ZNA,Dmitriev2026}, and to the possibility to realize them experimentally. In particular, the metallic framework of the 2d coordination polymers [\{Cu(bipn)\}$_3$Fe(CN)$_6$](ClO$_4$)$_{2}\cdot4$H$_2$O [bipn = bis(3-aminopropyl)-amine] \cite{Zhang2000} and [\{Cu(ept)\}$_3$Fe(CN)$_6$]{\allowbreak}(ClO$_4$)$_2\cdot5$H$_2$O [ept = N-(2-aminoethyl)-1,3-diaminopropane] \cite{Travnicek2001} adopts the geometry of the diamond-decorated honeycomb lattice. In both compounds, the Cu$^{2+}$ ions carry spin-$1/2$ moments, whereas the Fe$^{2+}$ ions are driven into a nonmagnetic low-spin ($S=0$) state by the strong ligand field of the cyanide groups. Although these materials are therefore not direct realizations of the present magnetic model, they demonstrate that the required lattice geometry is chemically accessible. This suggests that analogous coordination polymers in which the nonmagnetic hexacyanoferrate(II) units are replaced by paramagnetic divalent hexacyanometallate building blocks, while preserving the same framework topology, could provide a viable route toward realizing the magnetic diamond-decorated honeycomb lattice explored here. 

On the decorated honeycomb lattice, Kasteleyn physics appears because the low-energy sector maps onto an effective dimer model on the honeycomb lattice, with however a small and tunable density of monomers. Because the Kasteleyn transition arises from the strict global packing constraint that forces every lattice site to belong to a dimer, it is inherently fragile. Even a small finite density of monomer vacancies relaxes this constraint, allowing local rearrangements that smooth out the transition. Consequently, as proven by Heilmann and Lieb \cite{Lieb1970}, the free energy remains analytic everywhere in the physical monomer-dimer phase diagram except in the strict zero-monomer limit. One can therefore at best expect a crossover to take place (monomer-dimer models can only display genuine critical behavior in the form of Berezinskii-Kosterlitz-Thouless \cite{Otsuka2009,Lilichen}, Gaussian, or Potts-type crossovers \cite{Otsuka2011,Otsuka2013}). However, the density of monomers can be tuned to an arbitrarily small values, making the crossover increasingly sharp while shifting it towards lower temperatures. This provides an example of emergent Kasteleyn-type criticality and makes the present system a remarkable playground for studying the interplay between monomers and the Kasteleyn transition.

\paragraph{Heisenberg diamond-decorated honeycomb lattice}
\label{model}
\begin{figure}[t!]
\vspace{-0.2cm}
\centering
\begin{minipage}[b]{0.5\textwidth}
  \includegraphics[width=0.65\textwidth]{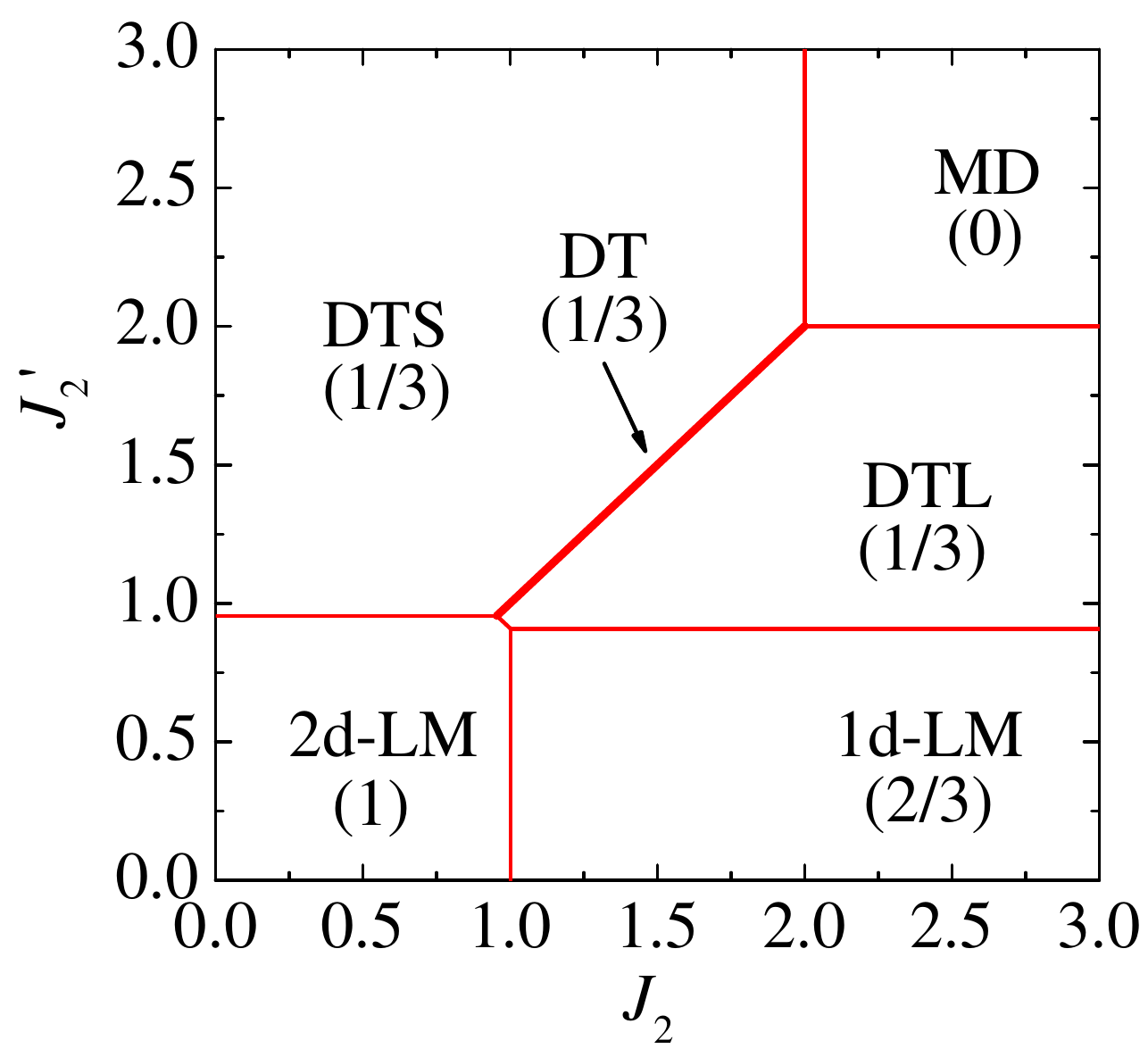}%
  \hspace{0.2cm}
  \raisebox{0.8cm}{\includegraphics[width=0.3\textwidth]{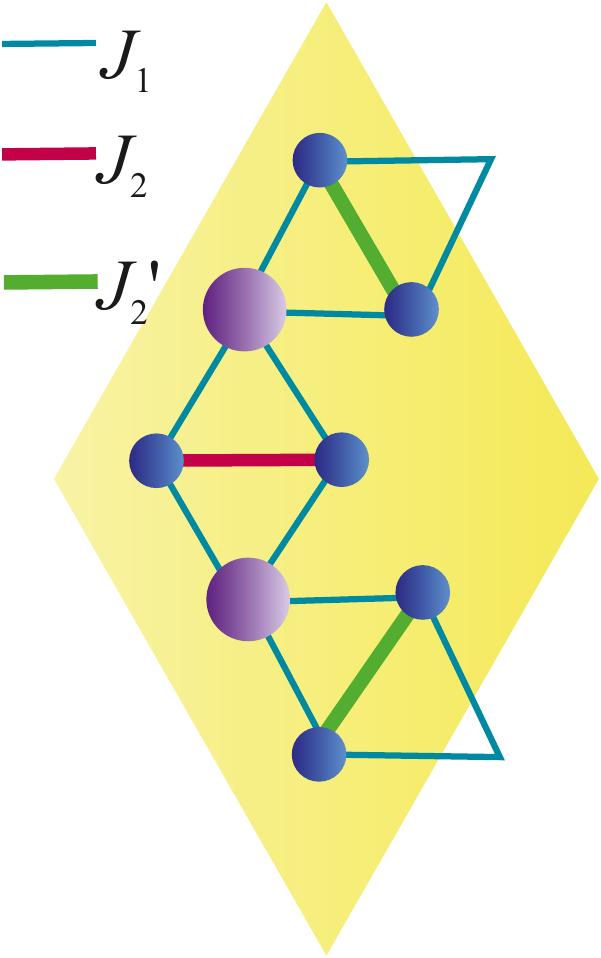}}
\end{minipage}%
\hspace{0.5cm}%
\begin{minipage}[b]{0.5\textwidth}
  \includegraphics[width=\textwidth]{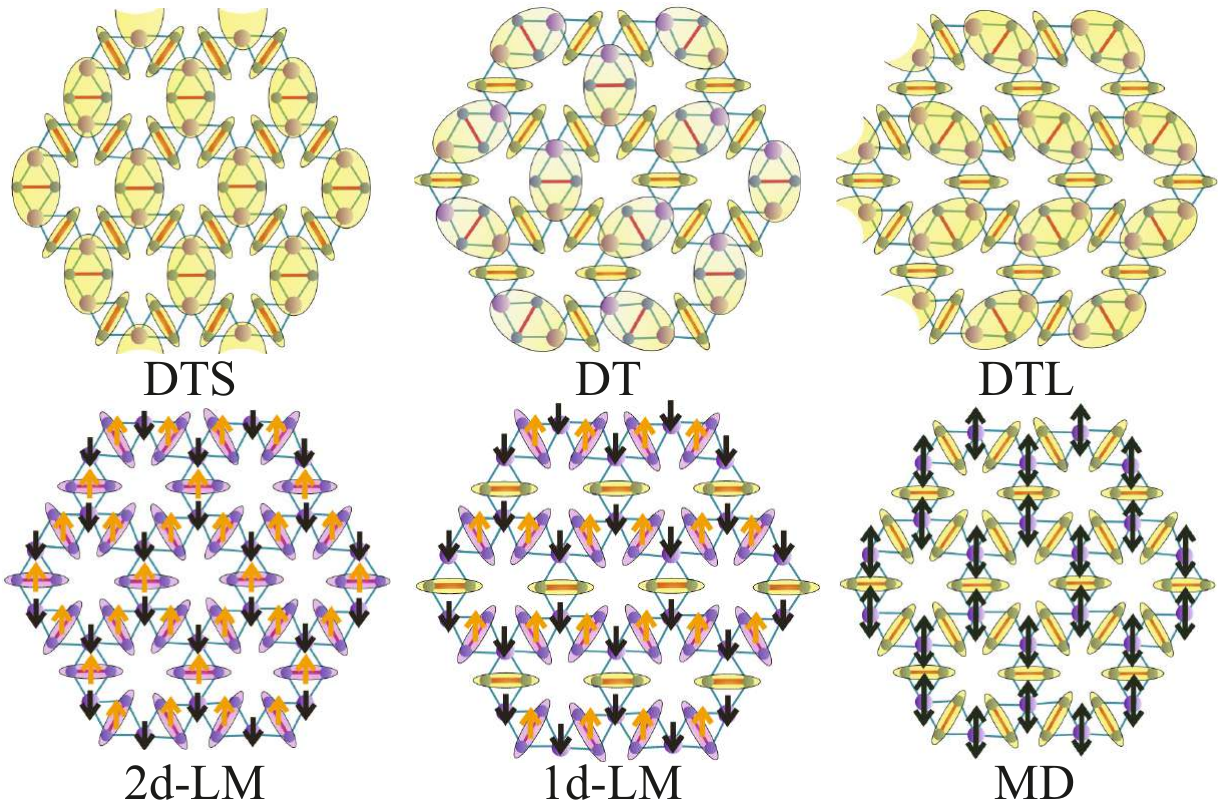}
\end{minipage}
\vspace{-0.5cm}
\caption{Upper panel: Ground-state phase diagram of the spin-$\tfrac{1}{2}$ Heisenberg antiferromagnet on the diamond-decorated honeycomb lattice in the $J_2$–$J_2'$ plane. Numbers in parentheses indicate the density of triplet dimers. The unit cell illustrates the definition of couplings $J_1$, $J_2$, and $J_2'$. Lower panel: Schematic illustrations of the six ground states: dimer-tetramer solid (DTS), dimer-tetramer (DT), dimer-tetramer liquid (DTL), 2d Lieb-Mattis (2d-LM), 1d Lieb-Mattis (1d-LM), and monomer-dimer (MD) phases.}
\label{fig:zfgspd}
\end{figure}

We consider two types of nearest-neighbor intra-diamond couplings $J_2$ and $J_2'$ (see Fig.~\ref{fig:zfgspd} and \textit{End Matter} for details). The other bonds of each diamond are all equal and denoted by $J_1=1$ hereafter serving as the energy unit. This implies that the total spin of each dimer bond is conserved, hence, each dimer is either in a singlet or triplet state, defining separate sectors in the Hilbert space.

We combine several complementary methods to analyze both ground-state and thermodynamic properties. These include exact diagonalization (ED), finite-temperature Lanczos method (FTLM) \cite{JaP:PRB94,prelovsek_bonca_2013,Wietek2019}, density matrix renormalization group (DMRG) \cite{White1992}, and sign-problem-free quantum Monte Carlo (QMC) \cite{Sandvik1991,Syljuasen2002,Sandvik2003,Stapmanns2018,Weber2022} simulations for the original model. For further implementation details, we refer to our previous studies on the square-lattice version of the model \cite{Caci2023,Karlova2024} and the Supplemental Material \cite{supp}. The effective dimer model (EDM) and the effective monomer-dimer model (EMDM) are numerically solved via the transfer-matrix \cite{Lieb1967,Grande2011} 
and corner transfer matrix renormalization group (CTMRG) methods \cite{Nishino96,ORUS14},
see \textit{End Matter} and Supplemental Material \cite{supp} for details.

Figure~\ref{fig:zfgspd} shows the ground-state phase diagram, as obtained using
the ALPS implementation of DMRG \cite{Bauer2011}
on a system of $4 \times 4$ unit cells ($N$=128 spins) under periodic boundary conditions. 
We first focus on the symmetric line $J_2=J_2'$.
For $J_2=J_2'<0.9548$, the system realizes the 2d-LM phase, where all dimers are triplets and the monomeric spins are predominantly oriented in the opposite direction. This phase resembles  classical ferrimagnetic order, but with a small quantum reduction of the magnetization. For $J_2=J_2'>2$, the system enters the MD phase, where all dimers are singlets separated by free (paramagnetic) monomeric spins.

In the intermediate region, $0.9548<J_2=J_2'<2$, the system enters a macroscopically degenerate DT phase \cite{Morita2016}, with singlet tetramers (large ovals in Fig.~\ref{fig:zfgspd}) separated by singlet dimers (small ovals).
Here, a singlet tetramer refers to a singlet formed from a \emph{triplet} on the intra-diamond dimer compensated by a \emph{triplet} formed from the two neighboring monomeric spins 1/2. 
The DT ground-state manifold can be mapped onto that of dimer coverings of the underlying honeycomb lattice, where each occupied bond corresponds to a singlet tetramer. This mapping implies
a ground-state entropy in a torus geometry of $0.06\cdot\ln 2$ per spin \cite{Wu2006,Morita2016,Wannier1950,Kasteleyn1963}.

To gain further insight into the undistorted case, we performed ED on a 32-site cluster comprising 12 dimers and 8 monomeric sites. In the DT phase, the ground state lies in the sector with exactly four triplet dimers, consistent with a triplet density of 1/3. The lowest excitations then correspond to neighboring sectors with three, five, and six triplet dimers, as well as the first excited state within the four-triplet sector (see \textit{End Matter} for the 
ED spectra). The excitation gap is maximized near $J_2 \approx 1.4$, identifying this region as the most stable realization of the DT phase and a natural choice for mapping onto the EDM. We will therefore focus 
on this parameter regime. 

\paragraph{Kasteleyn signatures in a weakly distorted version}
\label{results}
Next we address the effects of breaking the symmetry $J_2 = J_2'$. 
The macroscopically degenerate DT phase then splits into two distinct regimes, compare Fig.~\ref{fig:zfgspd}. The system either favors a non-degenerate DTS phase, which has an ordered pattern of singlet tetramers on all vertical bonds, or the DTL phase, where singlet tetramers form along diagonal bonds. This phase remains massively degenerate, but has no residual entropy, distinguishing it from typical highly frustrated states.

Furthermore,  
in the region $J_2>J_2'$, this spatial anisotropy leads to the emergence of a 1d-LM phase in which diamond chains with a zig-zag pattern are effectively decoupled from each other by dimer singlets on all $J_2$-bonds. This 1d-LM ferrimagnetic phase thus has 1d characteristics, distinct from the 2d-LM counterpart.
\begin{figure}[t!]
    \centering
        \includegraphics[width=0.4\textwidth]{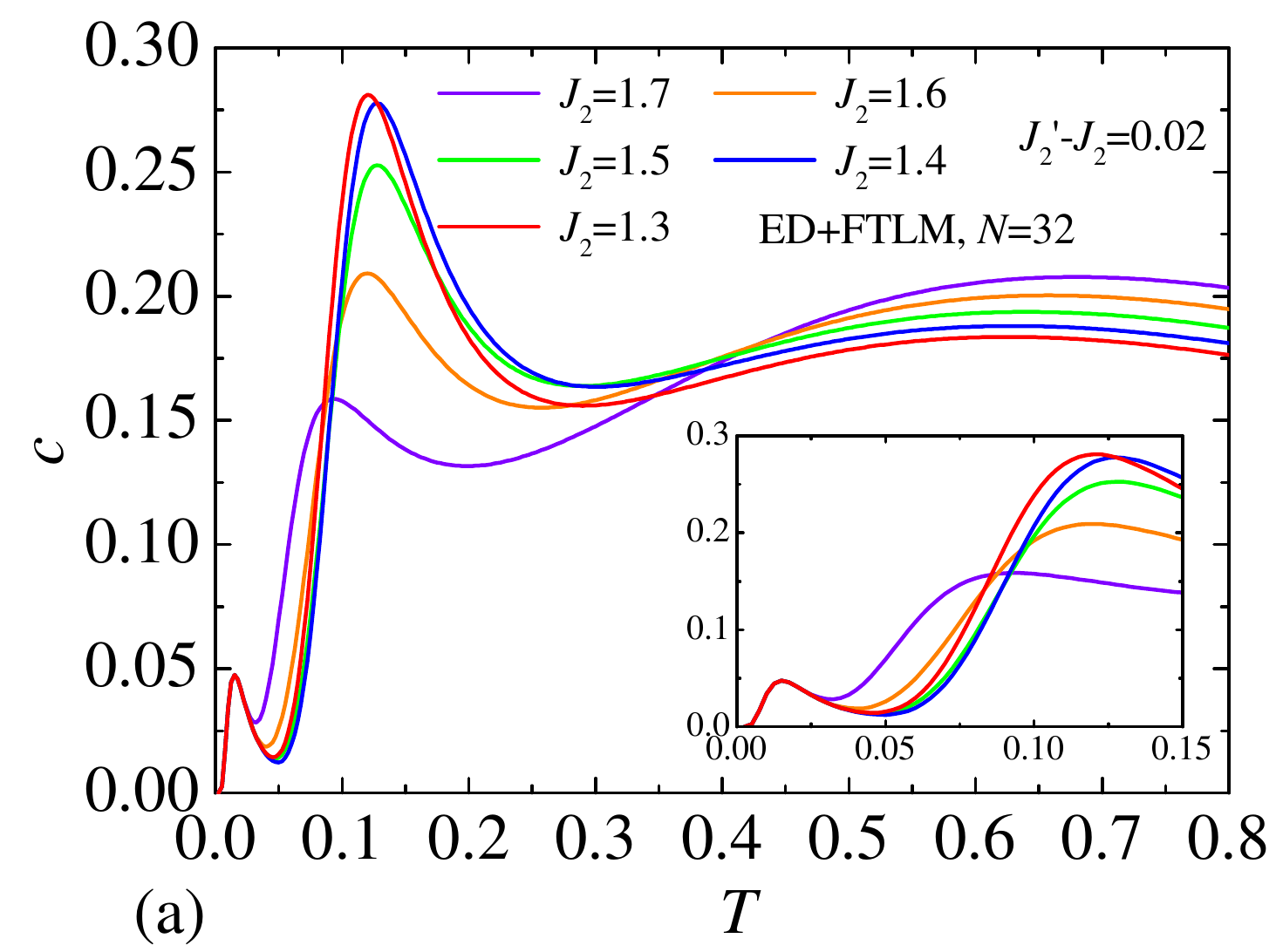} 
         \includegraphics[width=0.4\textwidth]{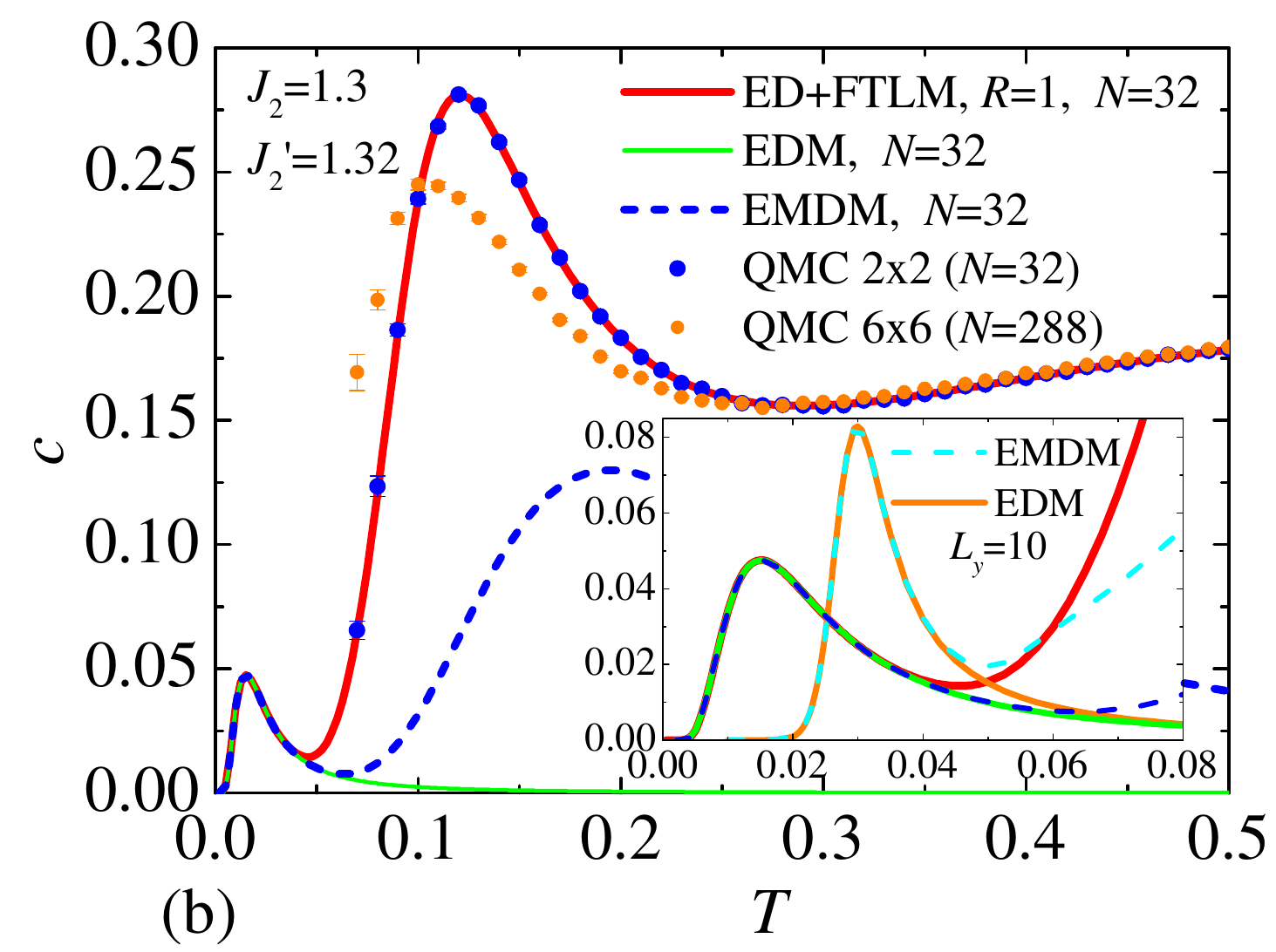}
				\vspace{-0.4cm}
\caption{Specific heat per spin $c$ as a function of temperature for the spin-$\frac{1}{2}$ Heisenberg model on the diamond-decorated honeycomb lattice with $J_2' - J_2 = 0.02$. (a) ED+FTLM results for $N = 32$ spins for several values of $J_2$. (b) Comparison of ED+FTLM results with QMC data at $J_2 = 1.3$, $J_2' = 1.32$ for two system sizes, $N = 32$ and $N = 288$. Inset: Comparison of ED+FTLM results with the effective dimer model (EDM) and the effective monomer--dimer model (EMDM) for $2 \times 2$ and $\infty \times 10$ unit cells.
}
\label{fig:speci}
\end{figure}

From now on, we focus on the regime where the DTS forms the ground state and the DTL appears as a low-energy excitation, where Kasteleyn physics is expected. To shed light on the thermal properties, we present in Fig.~\ref{fig:speci}(a) the specific heat obtained from ED combined with FTLM for $N = 32$ spins. For various values of $J_2$ and fixed $J_2' - J_2 = 0.02$, we observe three peaks at well separated temperatures ($T \approx$ 0.02, 0.1, and 0.7). To assess the validity of the ED+FTLM approach at higher temperatures and larger system sizes, we compare in Fig.~\ref{fig:speci}(b) the specific heat for $J_2 = 1.3$, $J_2' = 1.32$ with results from QMC simulations. Excellent agreement is found in the accessible temperature range. In particular, the broad intermediate peak is consistently reproduced across all methods, while the larger $N = 288$ QMC cluster reveals its progressive suppression with system size. The low-temperature peak is inaccessible to QMC due to sampling challenges in the dimer basis \cite{Stapmanns2018,Weber2022,Caci2023}, but remains clearly resolved by ED+FTLM. The results presented below provide strong evidence that the low-temperature peak, instead of reducing, sharpens with increasing system size [see the inset in Fig.\ref{fig:speci}(b)].

Let us now concentrate on the interpretation of this low-temperature peak. The proximity of the degenerate DT line suggests that it is essentially due to the lifting of degeneracy in the ground-state manifold of the DT phase due to the difference between $J_2$ and $J_2'$. If that is the case, the very same low-temperature peak should be present in a simple EDM description of the low-energy excitations with different energies on vertical and tilted bonds. This model is precisely the model for which Kasteleyn discovered the transition that bears his name \cite{Kasteleyn1963,Wu2006}. To extend this effective description to higher energies, we note that 
any singlet tetramer in the ground-state manifold of the DT phase can be replaced by a diamond with an intra-dimer singlet and two unpaired spins, {\it i.e.}, a pair of monomers (with an extra degeneracy due to the spin). In the effective description, monomers correspond to free spin-1/2 entities and therefore carry an additional twofold degeneracy compared to the classic monomer–dimer model. As in the pure dimer model on the honeycomb lattice, these monomers can be further separated by exchanging singlet dimers and singlet tetramers along the path. Note that all these states are still exact eigenstates. So the low-temperature properties of the present model should be identical to those of the EMDM on the honeycomb lattice with appropriate effective dimer and monomer energies. Details of this mapping are given in the \textit{End Matter} and in the Supplemental Material \cite{supp}.

To validate this interpretation, we compare in Fig.~\ref{fig:speci}(b) the specific heat for $J_2' - J_2 = 0.02$  with  
results for the EDM and EMDM on the same finite-size system. Both effective models indeed capture the low-temperature peak quantitatively. In particular, the EDM reproduces the ED+FTLM data up to $T \approx 0.04$, confirming that the low-temperature peak is dominated by the ground-state configurations of the DT phase. The EMDM further improves the agreement at higher temperatures and also captures the subsequent rise in the specific heat, showing that monomer excitations become relevant in this temperature range \footnote{Note, however, that for $J_2 = 1.3$, there are further excitations at energies below the monomer ones, see Fig.~\ref{fig:gap}, such that the EMDM no longer yields a quantitatively accurate description either
at these higher temperatures.
}. 
The transfer-matrix results for the EDM and EMDM with the geometry of a tube comprising an infinite number of unit cells in one direction and 10 unit cells ($L_y=10$) in the other reveal that the low-temperature peak of the specific heat increases steadily and undergoes a slight shift toward higher temperatures with increasing system size.

\begin{figure}
	\centering
	\includegraphics[width=0.4\textwidth]{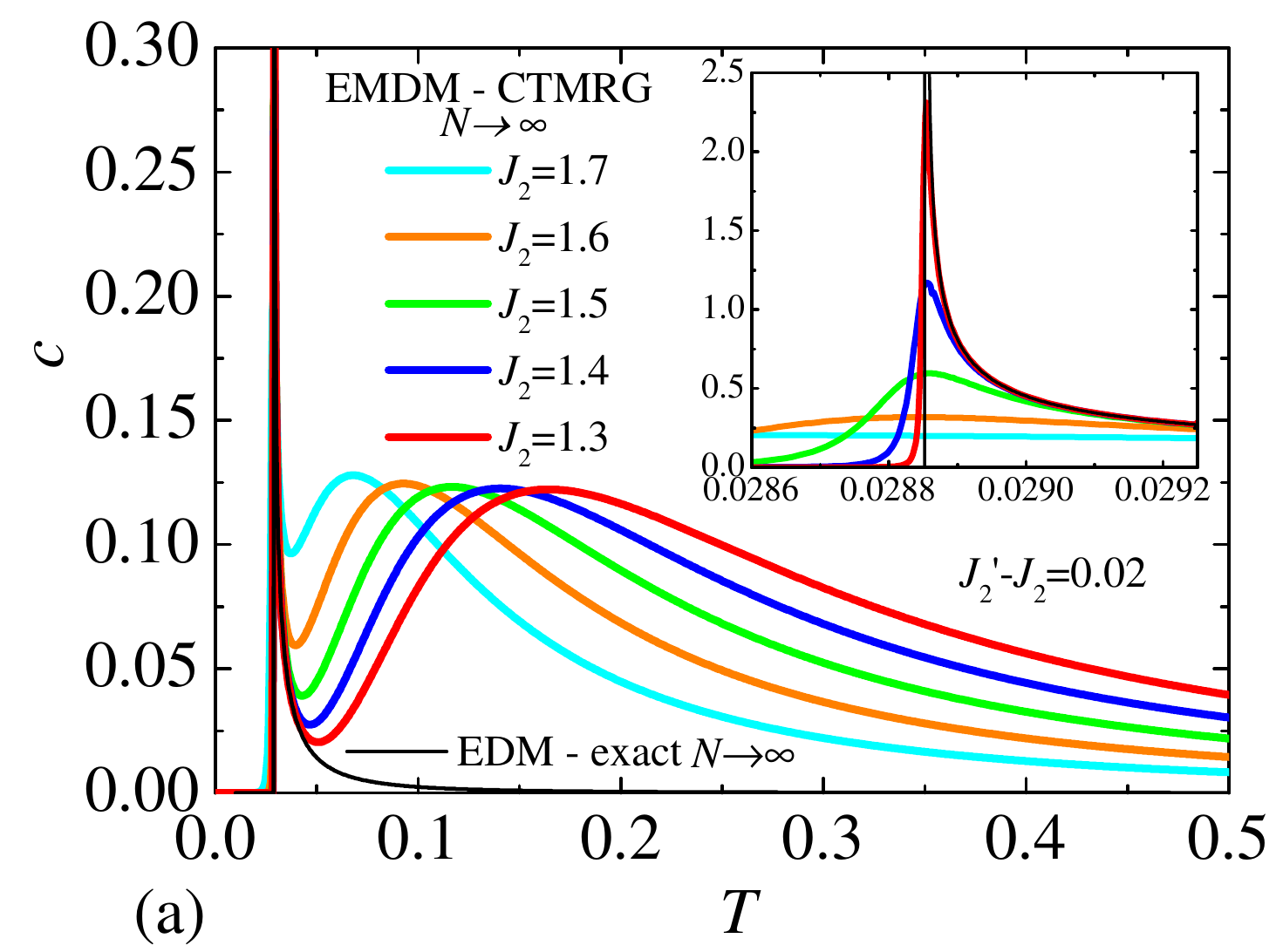}
	\includegraphics[width=0.4\textwidth]{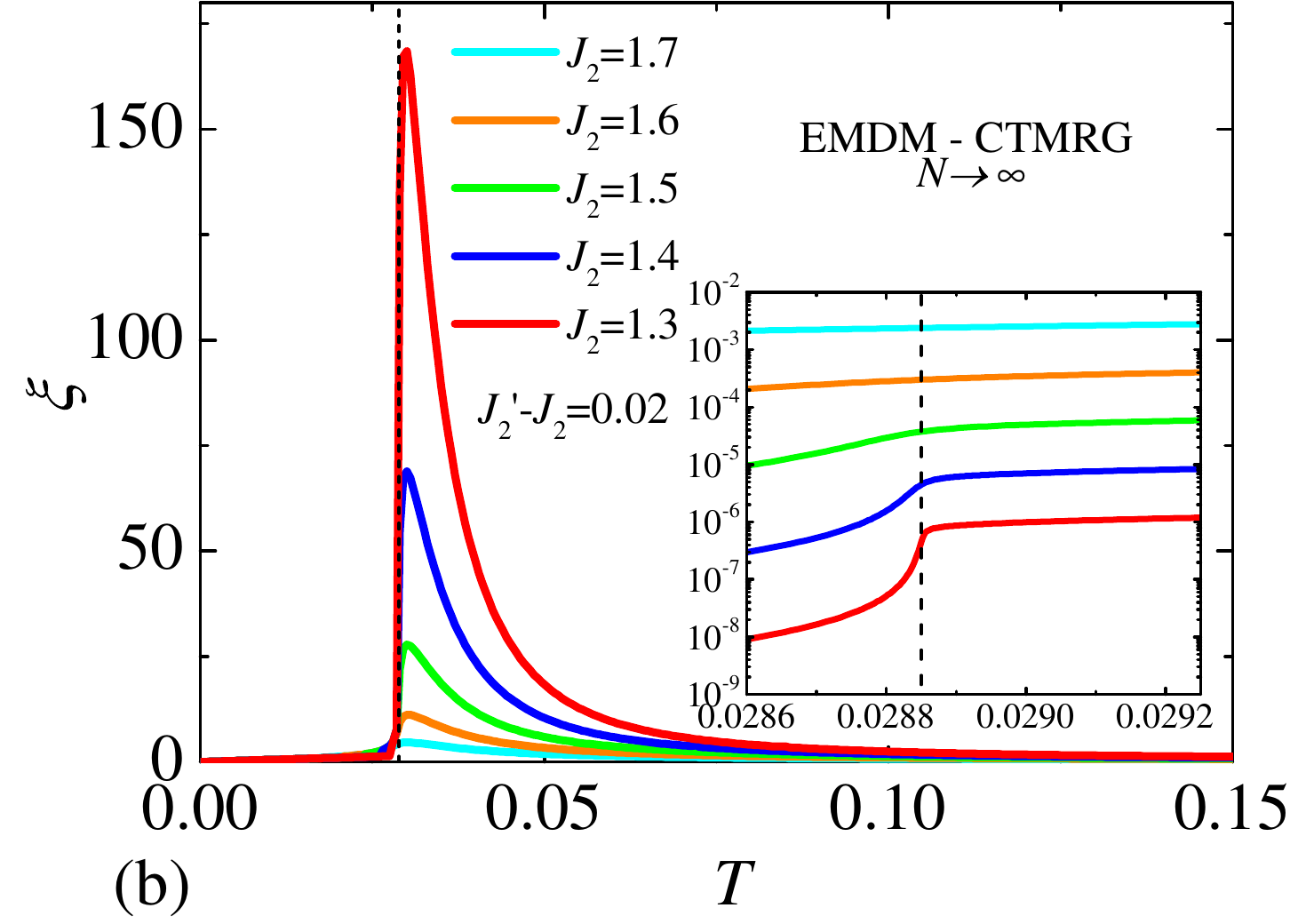}
	\vspace{-0.3cm}
		\caption{
(a) 
Specific heat $c$ and (b) correlation length $\xi$
for the spin-$\frac{1}{2}$ Heisenberg model on the diamond-decorated honeycomb lattice in the weakly distorted regime $J_2' - J_2 = 0.02$ as obtained from the EMDM using CTMRG in the thermodynamic limit for distinct values of $J_2$. Insets highlight the 
regime around $T_{\rm K}$:  in panel (a) for the specific heat and in panel (b) for the monomer density $n_M$.
In panel (a), the exact result for the specific heat of the EDM is also shown by a thin black line.}  \label{fig:ctmrg}
\end{figure}

\begin{figure}[t!]
	\centering
	\includegraphics[width=0.4\textwidth]{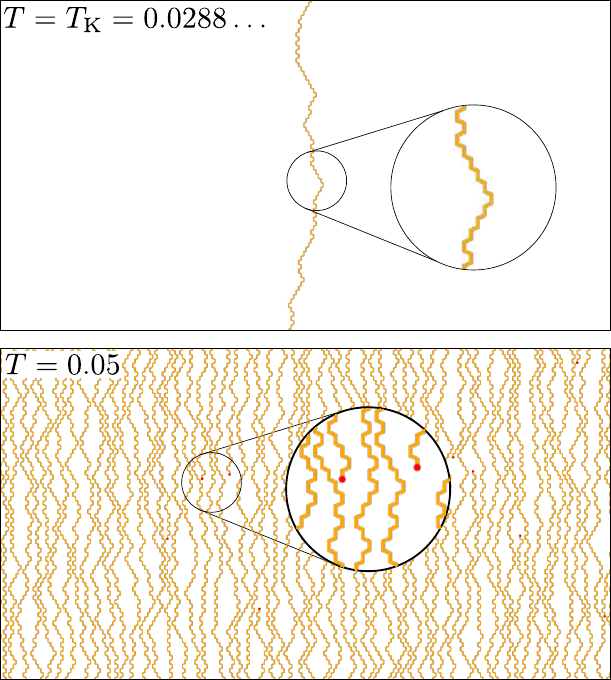}
	\caption{CTMRG snapshots of the EMDM for a cross section  
		of size $122\times 66$, $J_2=1.3$ and $J_2'-J_2=0.02$  at $T=T_{\rm K}=0.0288...$ (top) and $T$=0.05 (bottom). Orange lines denote dimers where the occupation differs from the ground state.
The resulting string-like excitations run mostly vertically.
Red dots denote monomers.
	\label{fig:snapmain}
}
\end{figure}

Having validated the effective description and illustrated its microscopic configurations, we now analyze the thermodynamic response of the EMDM using the CTMRG technique (see \textit{End Matter} for further information). This tensor-network method is well suited for 2d classical systems in the thermodynamic limit \cite{Nishino96,ORUS14}, and provides essentially exact results once the bond dimension is converged. Figure~\ref{fig:ctmrg}(a) shows the specific heat as a function of temperature for various values of $J_2$, revealing a sharp low-temperature peak associated with dimer physics, followed by a broader maximum due to monomer excitations. The inset highlights the peak structure around $T_\mathrm{K} \approx 0.0289$, where the specific heat of the EMDM closely follows the inverse square-root divergence $1/\sqrt{T-T_\mathrm{K}}$ of the pure EDM indicated by a thin black line (see Eq.~\eqref{eq:cv} in the \textit{End Matter}). In the limit of vanishing monomer density, the sharp low-temperature peak of the specific heat thus rigorously approaches the inverse square-root singularity of the hard-dimer problem 
with $T_\mathrm{K}=(J_2'-J_2)/\ln 2\approx 0.0289$ \cite{Kasteleyn1961}.

While the EMDM retains this thermodynamic signature, the finite monomer density $n_M$ ultimately smooths the transition into a 
round but extremely steep crossover. This is confirmed by the behavior of the correlation length $\xi$ shown in Fig.~\ref{fig:ctmrg}(b), which displays a pronounced maximum near $T_\mathrm{K}$ but remains finite across the full temperature range. The finite value of the correlation length $\xi$ is consistent with the absence of a true thermodynamic singularity in 2d monomer–dimer models in accordance with the theorem of Heilmann and Lieb \cite{Lieb1970}. However, the proliferation of string excitations as a hallmark of the Kasteleyn transition is still the key mechanism for the crossover from the DTS to the DTL regime, as evidenced by the snapshots 
in Fig.~\ref{fig:snapmain}. 
Well below $T_{\rm K}$ the system is nearly string-free, consistent with a frozen DTS. At the Kasteleyn temperature $T_{\rm K}$ only isolated strings are present, whereas slightly above it a dense network of strings rapidly develops.
Additional snapshots covering a broader range of parameters are presented in Fig.~\ref{fig:snap} of the \textit{End Matter}.

The monomer density is exponentially small at low temperatures, and even if it exhibits a rapid increase around $T_\mathrm{K}$, it remains extremely small; for instance, it is of order $10^{-6}$ around $T_\mathrm{K}$ for $J_2=1.3$ and $J_2' - J_2=0.02$ (see inset of Fig.~\ref{fig:ctmrg}(b)). This sharp but low-amplitude rise is linked to the proliferation of monomer-induced strings (see snapshots in Fig.~\ref{fig:snap} of the \textit{End Matter}) and is responsible for the broadening of thermodynamic features.
By tuning $J_2' - J_2$, the monomer density can be reduced to arbitrarily small values, allowing the system to asymptotically approach the true Kasteleyn critical point. This makes the spatially anisotropic diamond-decorated Heisenberg model a rare and tunable quantum platform for realizing emergent classical criticality driven by dimer constraints
in a frustrated quantum magnet.

\paragraph{Conclusion and Perspectives}
\label{conclusion}
Macroscopic ground-state degeneracy  is a hallmark of highly frustrated magnets, at least at the classical level \cite{Lacroix2011,Diep2020}.
The present model provides an example where this degeneracy is not lifted even by quantum fluctuations.
Nevertheless, according to the third law of thermodynamics \cite{Nernst1906}, one expects nature to lift this
ground-state degeneracy in experimental systems. A lattice distortion that renders the coupling constants
spatially anisotropic, {\it i.e.}, the situation studied in this Letter, is one mechanism to lift such a degeneracy.
Generically, one expects a weak splitting of a degenerate ground-state manifold to give rise to a low-temperature
peak in the specific heat. The present example demonstrates further that such a degeneracy lifting can give rise to  emergent Kasteleyn physics. The examples of the staggered diamond-decorated square lattice and the anisotropic fully frustrated kagome bilayer \cite{Yaremchuk2025}
presented in the Supplemental Material \cite{supp} suggest that emergent Kasteleyn physics is a generic feature of a broader class of frustrated quantum magnets with emergent dimer degrees of freedom. Our work thus provides a bridge between quantum magnetism and the statistical mechanics of classical dimer models, opening new avenues for the study of topological and geometrically constrained phases in frustrated quantum spin systems.

\begin{acknowledgments}
We thank Lukas Weber for making his QMC code available for this study. 
We acknowledge funding from the European Union’s Horizon 2020 research and innovation programme under the Marie Skłodowska-Curie Grant Agreement No.~945380, the Štefánik program for Slovak–French bilateral collaboration (Grant Nos.~SK-FR-22-0011/49880PG and SK-FR-24-0005/53633XM), the EU NextGenerationEU through the Recovery and Resilience Plan for Slovakia under the project No.\ 09I03-03-V04-00403, the Swiss National Science Foundation (Grant No.~212082), the Slovak Research and Development Agency (Contract No.~APVV-24-0091), the National Research Foundation of Ukraine (Grant No.~2023.03/0063), and the French National Research Agency ANR (Grant Nos.~ANR-22-CE30-0042 and ANR-21-CE30-0033). Computing time for the QMC and ED simulations was provided by the IT Center at RWTH Aachen University and on the “osaka” cluster at the Centre de Calcul (CDC) of CY Cergy Paris Université, respectively.
\end{acknowledgments}

\bibliography{monomer-dimer}

\clearpage

\appendix
\section{End Matter}
\begin{figure}[t!]
	\centering
	\includegraphics[width=0.26\textwidth]{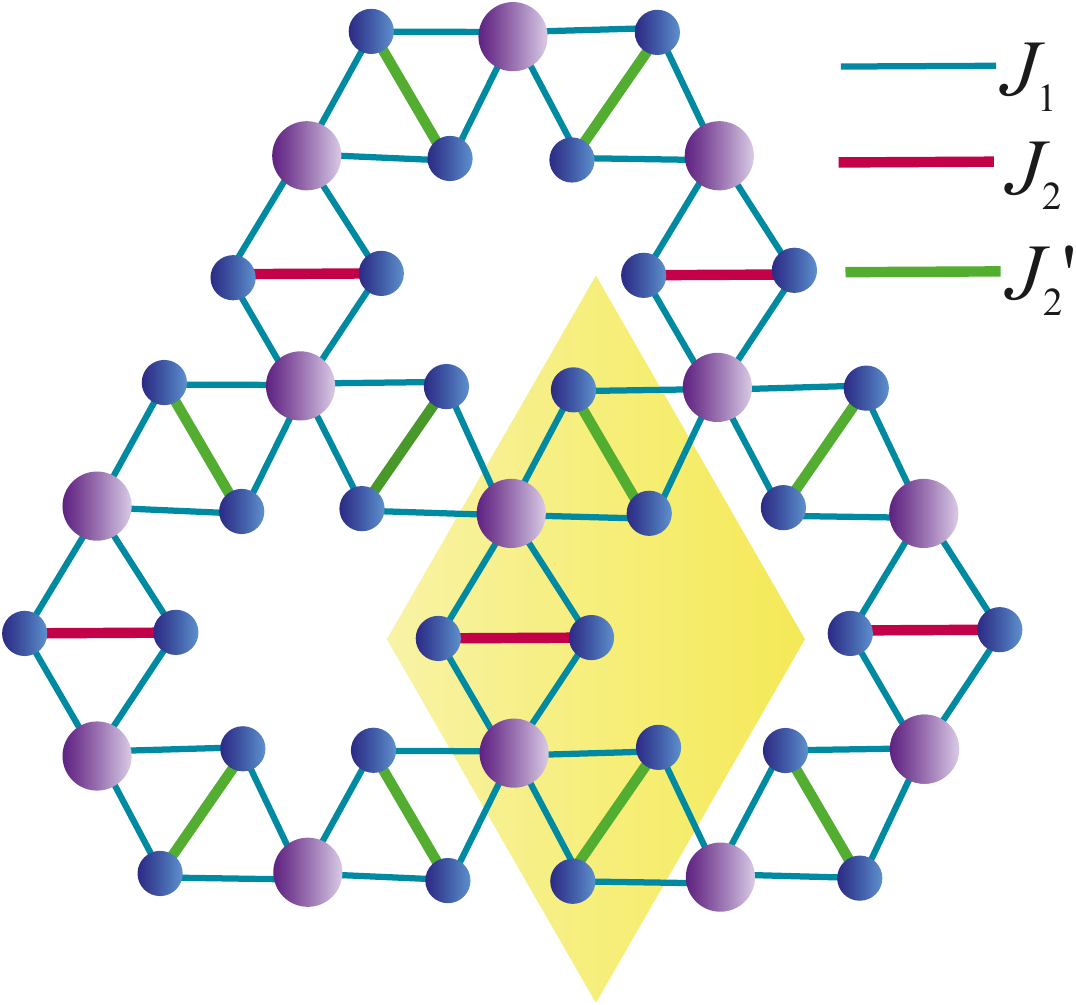}
	\vspace{-0.4cm}
	\caption{Part of the diamond-decorated honeycomb lattice with the unit cell shown in the yellow rhombus. The intra-dimer couplings $J_2$ and $J_2'$ within vertical and zig-zag diamonds are shown as thick red and green lines, whereas the monomer-dimer couplings $J_1$ are depicted as thin blue lines.}
	\label{fig:model}
\end{figure}

\paragraph{Hamiltonian}
\label{hamiltonian}
We consider the spin-$\frac{1}{2}$ Heisenberg model on a diamond-decorated honeycomb lattice, whose structure is illustrated in Fig.~\ref{fig:model} and which can be defined by the following Hamiltonian:
\begin{equation}
\hat{\cal{H}}=\sum_{\langle i,j \rangle}J_{i,j}\hat{\bm{S}}_i\cdot\hat{\bm{S}}_j,
 \label{eq:ham}
\end{equation}
where $\hat{\bm{S}}_i$ is a spin-$\frac{1}{2}$ operator located at the lattice site $i$, and $J_{i,j}$ denotes three different nearest-neighbor couplings $J_1, J_2$, and $J_2'$ 
as shown in Fig.~\ref{fig:model}. 
We employ periodic boundary conditions throughout.

The total spin of each dimer is a conserved quantity of this model. As a consequence, the Hilbert space fractionalizes into $2^{N_d}$
 independent sectors ($N_d$ denotes the number of dimers), which can be labeled according to whether each dimer is in its singlet or triplet state. This conservation law underlies the classification of phases and low-energy excitations of the model.

\begin{figure}[t!]
	\centering
	\includegraphics[width=0.85\columnwidth]{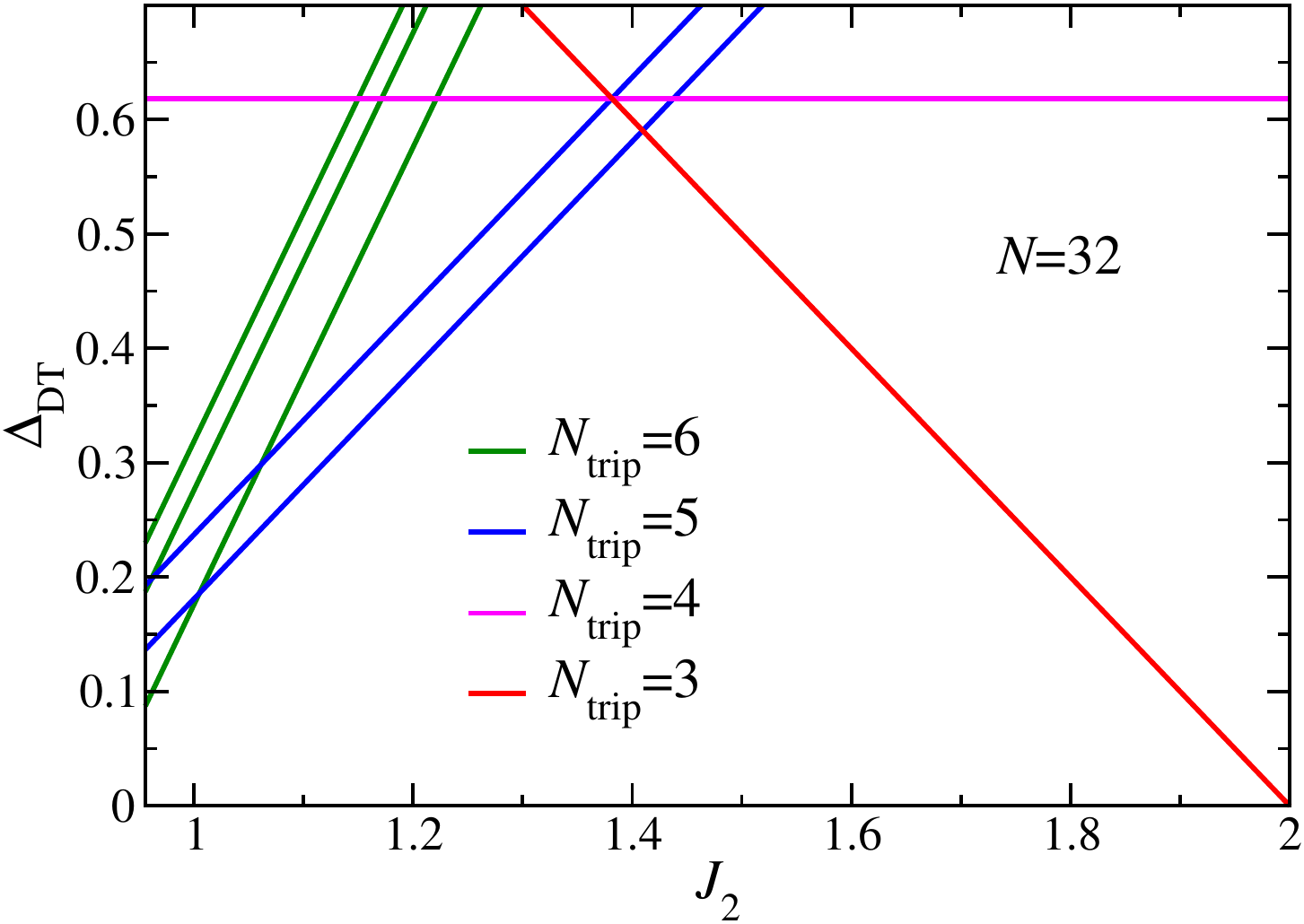}
\vspace{-0.5cm}
	\caption{Lowest excitation gaps $\Delta_{\rm DT}$ in the DT phase obtained from ED for a $2 \times 2$ system in the isotropic case as a function of $J_2=J_2'$.
	}
	\label{fig:gap}
\end{figure}

\paragraph{Low-energy gaps from ED in the undistorted case}
Figure~\ref{fig:gap} presents the lowest excitation gaps $\Delta_{\rm DT}$ as a function of $J_2$ in the DT phase for a $2 \times 2$ system with the total number of spins $N = 32$.  This system size has four triplet dimers $N_{\rm trip}=4$ within the DT phase.  For $J_2 \gtrsim 1.4$, the system is excited into a state with only $N_{\rm trip}=3$ and $\Delta_{\rm DT} = 2 J_1-J_2$ in this region. This is consistent with the 
MD phase emerging at $J_2 > 2$, where $N_{\rm trip}=0$.

 On the other hand, approaching the 2d-LM phase for $J_2 \lesssim 1.4$ the lowest-energy excitations involve progressively larger numbers of triplet dimers, reaching a maximum of 12 triplet dimers for $J_2 < 0.9559$ on the $2 \times 2$ system. 
 At $J_2 \approx 1.41$, the excitation gap is maximized at approximately $\Delta_{\rm DT} \approx 0.59$. This suggests that $J_2 \approx 1.4$ is an optimal region where the DT phase is most stable, making it a natural choice for emerging EDMs at low temperatures. Just above, {\it i.e.}, at $\Delta_{\rm DT} \approx 0.62$, there is another excitation with $N_{\rm trip}=4$. All these excitations are highly degenerate, namely $72$, $192$, and $88$-fold for the lowest $N_{\rm trip}=5$, $4$, and $3$ excitations, respectively, as compared to the $9$ ground states in the DT phase for the $2 \times 2$ system. Despite these relatively high excitation energies, their large degeneracies make their Boltzmann weights comparable to the total Boltzmann weight of the ground-state manifold already at $T \approx 0.2$. Therefore, a description in terms of only the classical ground states is expected to be valid only for $T \ll 0.2$.

\paragraph{Tensor-network formulation and CTMRG implementation}
\label{CTMRG}
We represent the partition function of the EMDM as a tensor network on the square lattice obtained by regrouping two vertical sites of the original honeycomb lattice.  Each local tensor encodes the hard constraint that every site is touched by either exactly one dimer or by a single monomer. This rule uniquely selects ten allowed local configurations for each tensor, each assigned the appropriate Boltzmann weight.  

We contract the infinite 2d network using the CTMRG method \cite{Nishino96,ORUS14}, which provides quasi-exact results for the thermodynamic limit once the bond dimension $\chi$ is converged. 
In our calculations we employed $\chi$ up to 320, which was sufficient to converge all observables reported in the main text.

\begin{figure*}[t!]
    \centering
		\includegraphics[width=\linewidth]{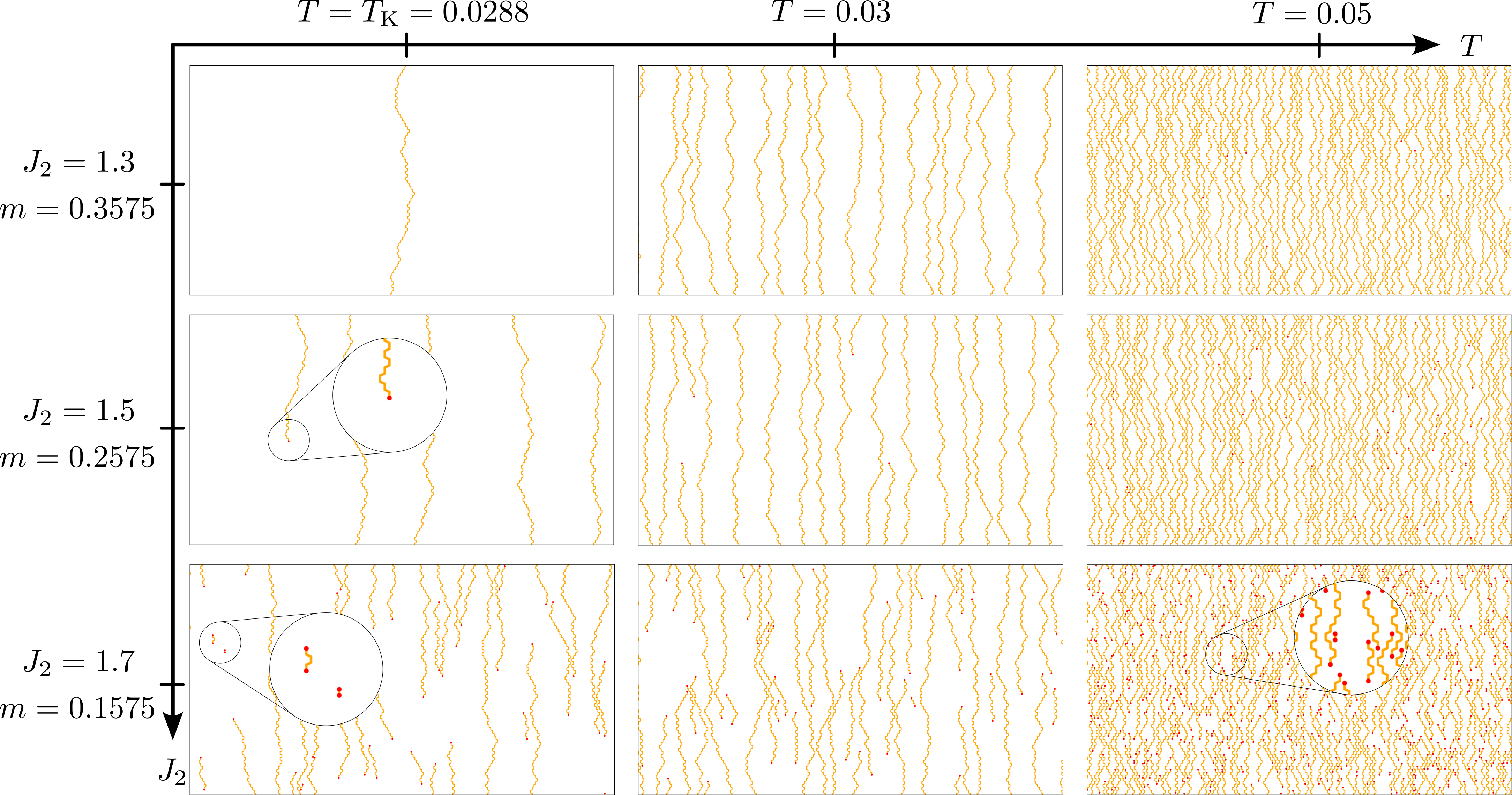}
		\caption{CTMRG snapshots of the effective monomer–dimer model (cross section  
of size $122\times 66$) for $J_2'-J_2=0.02$, shown at three distinct values of $J_2$ and temperatures. Orange lines correspond to string-like excitations and red dots denote monomers.
}	
    \label{fig:snap}
\end{figure*}

\paragraph{Mapping to the effective models}
\label{EMDM}
The EMDM takes into account only the lowest-energy localized states of singlet dimers and tetramers. Setting the energy of the
degenerate MD 
states to zero, we approximate the low-temperature partition function as follows
\begin{equation}
	\label{eq:Z_mdm}
	Z = \sum_{\cal{C}} {\rm e}^{- N_T \Delta'/T} {\rm e}^{-N_V \Delta/T} 2^{N_M},
\end{equation}
which describes a system of hard dimers and monomers on the honeycomb lattice. Here, $N_T$ ($N_V$) is the number of singlet tetramers on the tilted (vertical) diamond plaquettes, which are identified with the dimers on the corresponding bonds of the honeycomb lattice. The dimer energies $\Delta=J_2-2J_1$ and $\Delta'=J'_2-2J_1$ correspond to the excitations from the singlet dimer to the singlet tetramer on diamond plaquettes.
Note that $\Delta=-\Delta_{\rm DT}$ from above  such that $\Delta < 0$ in the DT phase. 
$N_M$ denotes the number of monomers, {\it i.e.}, free spins that are not enclosed by dimers (singlet tetramers). 
$\sum_{\cal{C}}$ runs over all possible configurations of dimers and monomers on the honeycomb lattice, where the following condition should be satisfied $2(N_T + N_V) + N_M = 2N_c$ ($N_c$ is the number of unit cells).
The critical properties of the model depend on
\begin{equation}
\delta=J'_2-J_2 .
\end{equation}
This becomes evident when we factor out the activity of vertical bonds:
\begin{equation}
	\label{eq:Z_mdm2}
	Z =  {\rm e}^{-N_c \Delta/T}\sum_{\cal{C}} {\rm e}^{- N_T \delta/T} {\rm e}^{-N_M m/T}.
\end{equation}
Here we introduced the effective free energy of monomers $m= -\Delta/2 -T\ln 2$.
We note that $m>0$ when $T<-\Delta/(2\ln 2)$. In this case the monomer activity is less than one, and it decreases exponentially with the temperature leading to the vanishingly small monomer density near $T_\mathrm{K}$ (see the inset of Fig.~\ref{fig:ctmrg}b).

\paragraph{Exact results for the pure dimer model}
\label{EDM}
In the limit of infinite effective monomer free energy ($m \to \infty$), our model reduces to the exactly solved hard-dimer problem \cite{Kasteleyn1963}. 
The specific heat per site is then given by \cite{nagle1989}
\begin{equation}
c_V (T) = 
\begin{cases}
\frac{\delta^2}{2 \pi T^2} \frac{{\rm e}^{\delta/T}}{\sqrt{1-\frac{1}{4}{\rm e}^{2\delta/T}}}, &
T \geq \frac{\delta}{\ln 2}, \\
0,	& 
 T < \frac{\delta}{\ln 2},
\end{cases}
\label{eq:cv}
\end{equation}
which exhibits an inverse square-root singularity when $T_\mathrm{K}$ is approached from above,
\begin{equation}
T_{\rm K} = \frac{\delta}{\ln 2} \simeq 1.4427\, \delta.
\end{equation}
The density of string excitations in this limit is
\begin{equation}
n_S(T) = 
\begin{cases}
\frac{2}{\pi} \arccos \left(\frac{1}{2}  {\rm e}^{\delta/T} \right), 
& T \geq \frac{\delta}{\ln 2}, \\
0,	 
& T < \frac{\delta}{\ln 2}.
\end{cases}
\label{eq:strings}
\end{equation}

\paragraph{Sampling of CTMRG snapshots}
\label{snapshots}
To visualize the nature of excitations across different temperature regimes, we show representative CTMRG snapshots of the EMDM in a finite $122\times 66$ region sampled from the Boltzmann distribution while averaging over all possible environments outside the central region \cite{uedaSnapshotObservation2D2005,rufino2025topologicaldevilsstaircaseconstrained}. Representative configurations are shown in Fig.~\ref{fig:snap}; the first and last panels of the top row of Fig.~\ref{fig:snap} are reproduced in Fig.~\ref{fig:snapmain} with additional zooms. In agreement with the dependence of the monomer energy on $J_2$, snapshots at larger $J_2$ display a higher density of monomer defects already at low temperatures. Around $T \simeq T_{\rm K}$, isolated strings proliferate, as expected for the Kasteleyn mechanism, but in the presence of monomers these strings are terminated into finite segments, turning the transition into a sharp crossover. With increasing temperature (or larger $J_2$), the monomer density rises further, shortening the average string length and generating a dense network of finite strings.

\end{document}



\title{\Large Supplemental Material for ``Approaching Kasteleyn transition in frustrated quantum Heisenberg antiferromagnets''
\\
\phantom{x}
}

\begin{abstract}
Here we provide details on ED, FTLM, CTMRG, and the transfer-matrix methods.
We 
also explain
how to map the Heisenberg model on the non-uniform diamond-decorated  lattices onto the effective monomer-dimer models
and provide complementary results for the diamond-decorated square lattice.
Finally, we argue that an anisotropic version of a fully frustrated kagome bilayer can be expected to also exhibit emergent Kasteleyn physics.
\end{abstract}

\date{\today}

\maketitle


\tableofcontents

\section{Exact diagonalization (ED)}

To diagonalize the original spin-1/2 Heisenberg model, we follow the same strategy as in previous investigations
of the square-lattice version of the model \cite{Caci2023,Karlova2024}. We start by rewriting the Hamiltonian as a composite spin model where each dimer gives rise either to a spin 0 or a spin 1. Next, we enumerate the topologically distinct
configurations of dimer spins 0 and 1. For the $2 \times 2$ diamond-decorated honeycomb lattice
($N=32$ spins 1/2), we find 96 topologically distinct configurations with multiplicities ranging from 1 to 168 for $J_2'=J_2$.
For evaluation purposes of the case $J_2' \ne J_2$, we need to distinguish primed and unprimed triplet dimers, thus increasing the number of distinct configurations to 232 and reducing the maximal multiplicity to 80; nevertheless, the spectra for a given configuration remain the same as in the case $J_2'=J_2$. For each of the 96 topologically distinct configurations, we then use $S^z$ conservation, aspects of SU(2) symmetry, and possibly spatial symmetries. Subsequently, we attempt
a full diagonalization in each sector. In this manner, we obtain 39\%\ of the spectrum (corresponding to approx.\ 96\%\ of the total entropy of the system). In particular, we have the full spectra in all sectors with $N_{\rm trip}\le8$, going well beyond the dimer-tetramer (DT) phase of interest for the present purposes.
For the sectors that we are not able to diagonalize completely, we use the
finite-temperature Lanczos method (FTLM) \cite{JaP:PRB94,prelovsek_bonca_2013,Wietek2019}.
Finally, we compute the thermodynamic quantities for the desired values of $J_2'$ and $J_2$ using post-processing of the full and Lanczos spectra previously obtained for all possible dimer patterns.

It turns out that  a single Lanczos run per sector suffices to obtain results where the effect of the random sampling inherent to FTLM is no longer visible in the combined ``ED+FLTM'' results.
This can be attributed to those sectors lying high in energy
that are treated by the approximate FTLM method,
thus becoming relevant only at elevated temperatures while the low-temperature behavior
arises from the sectors that are treated by exact full diagonalization,
in combination of the FTLM method becoming exact in the
high-temperature limit for large Hilbert space dimensions (see, e.g., Ref.~\cite{Wietek2019} and references therein).

For the square lattice, we can reuse the
exact diagonalization (ED) results for $N=40$ spins from our previous work \cite{Karlova2024} thanks to
the $J_2$ bonds yielding an effective chemical potential on the dimers;
we just need to distinguish $J_2$ and $J_2'$ dimers in the postprocessing step.
In addition, here we complement the previous results again by FTLM for those sectors that are too
large to be diagonalized completely.
The DT phase is still included in the fully diagonalized
range $N_{\rm trip.}\le 8$ \cite{Karlova2024}
even if in the $N=40$ square-lattice case full diagonalization is possible only for a smaller fraction of the
Hilbert space than for the $N=32$ honeycomb lattice. Consequently, a single Lanczos run suffices again in those
sectors that are too large to be diagonalized completely (dimension $\gtrsim 200\,000$) to obtain 
essentially exact ``ED+FTLM'' results also for the square-lattice version of the model at all temperatures
throughout the DT phase.

\section{
Diamond-decorated square lattice: effective monomer-dimer model} 
\label{Square}

	\begin{figure*}
  \centering
\includegraphics[width=0.3\textwidth]{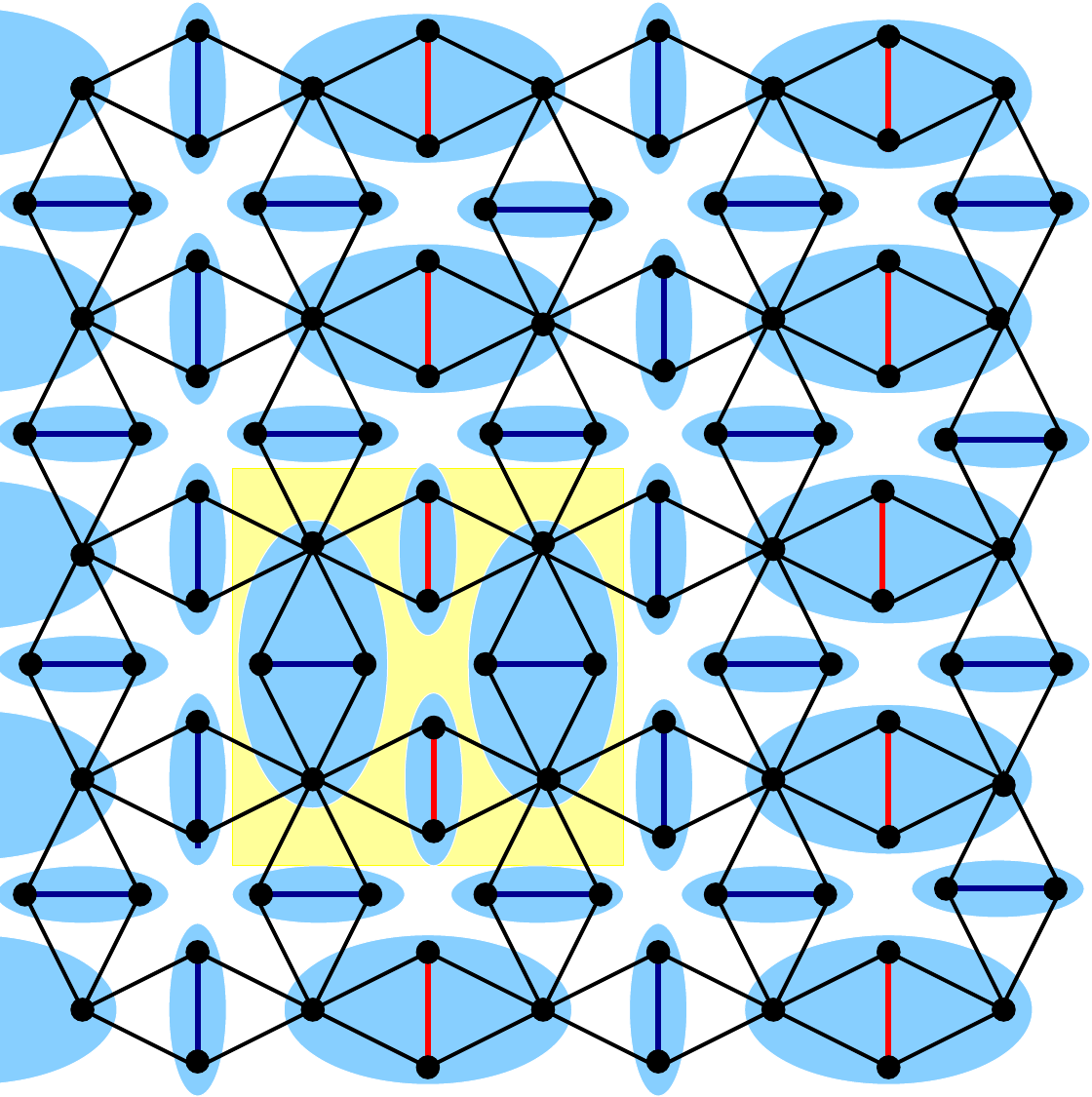}
\vspace*{0.3cm}
\includegraphics[width=0.3\textwidth]{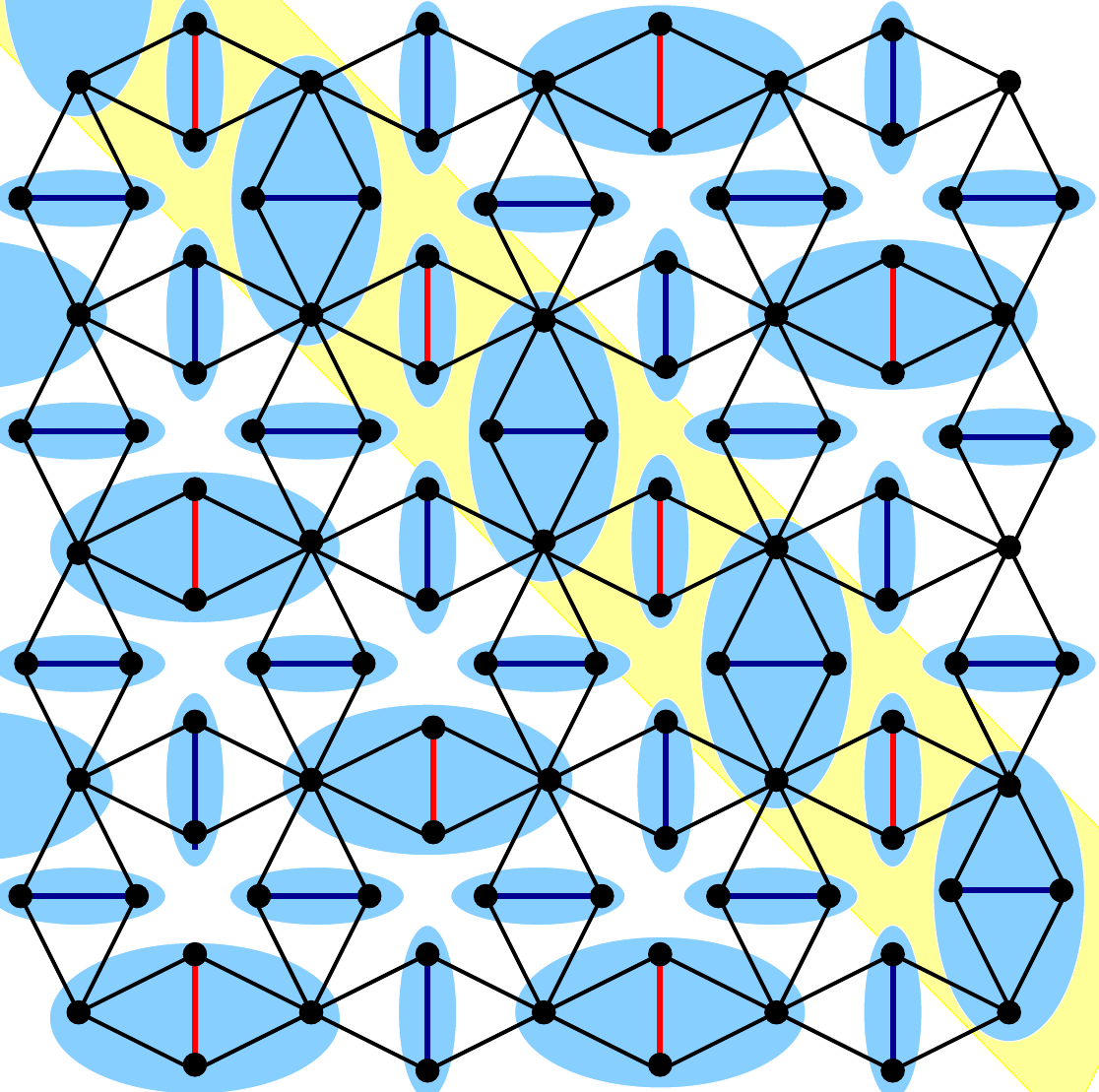}
\hspace*{0.3cm}
\includegraphics[width=0.3\textwidth]{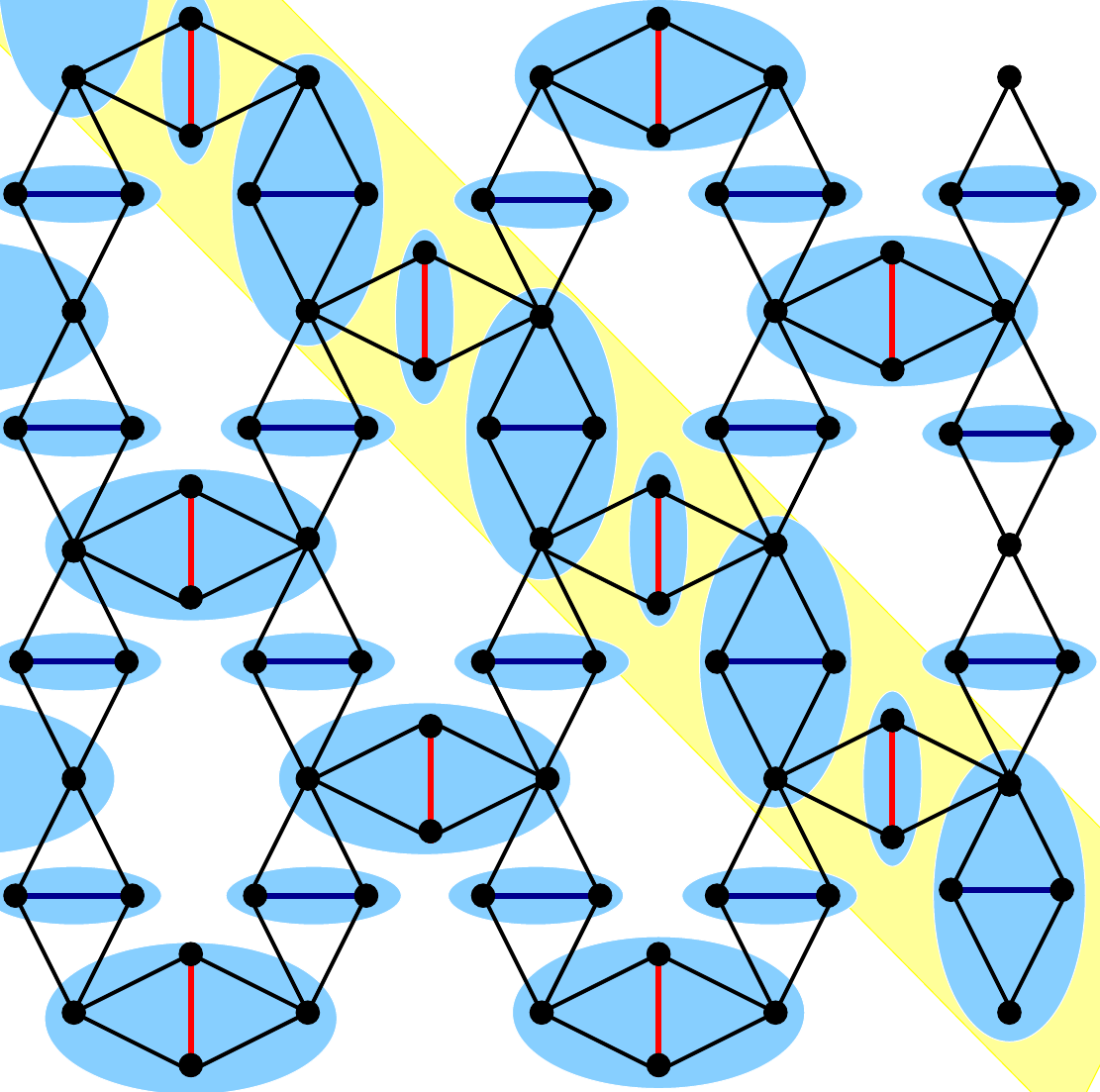} \\
\includegraphics[width=0.26\textwidth]{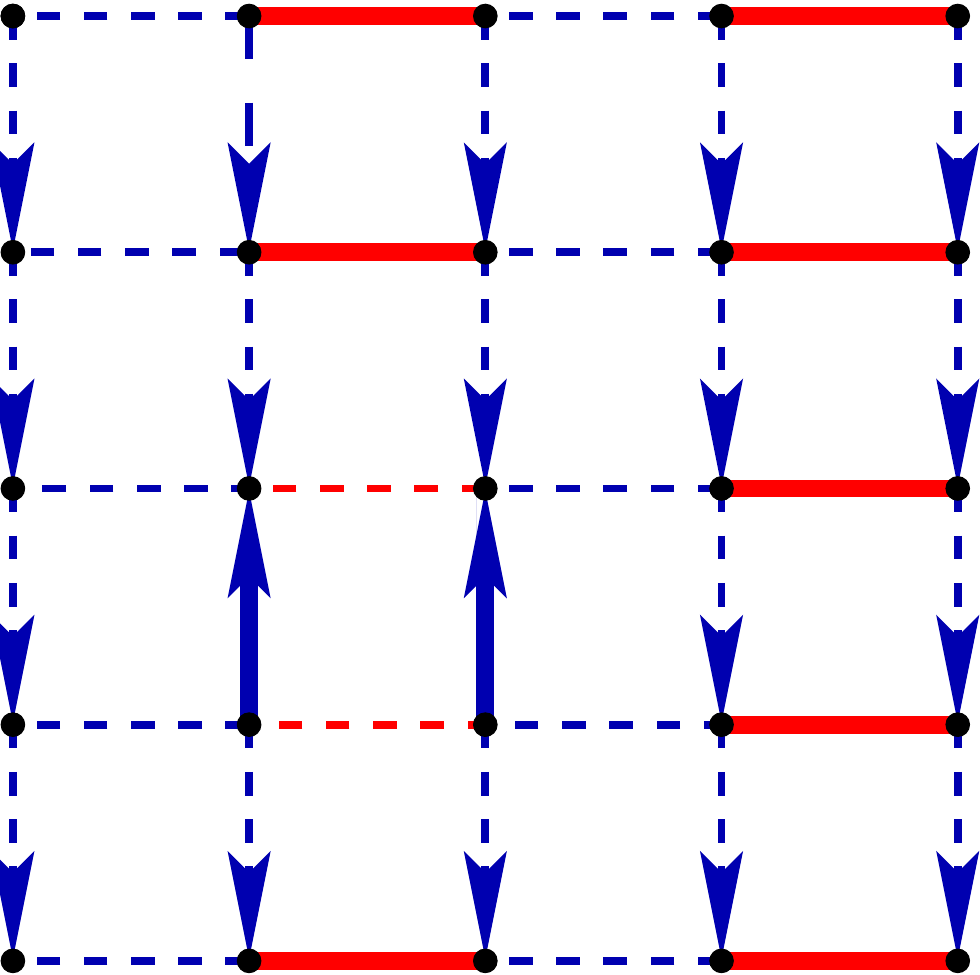}
\hspace{0.7cm}
\includegraphics[width=0.26\textwidth]{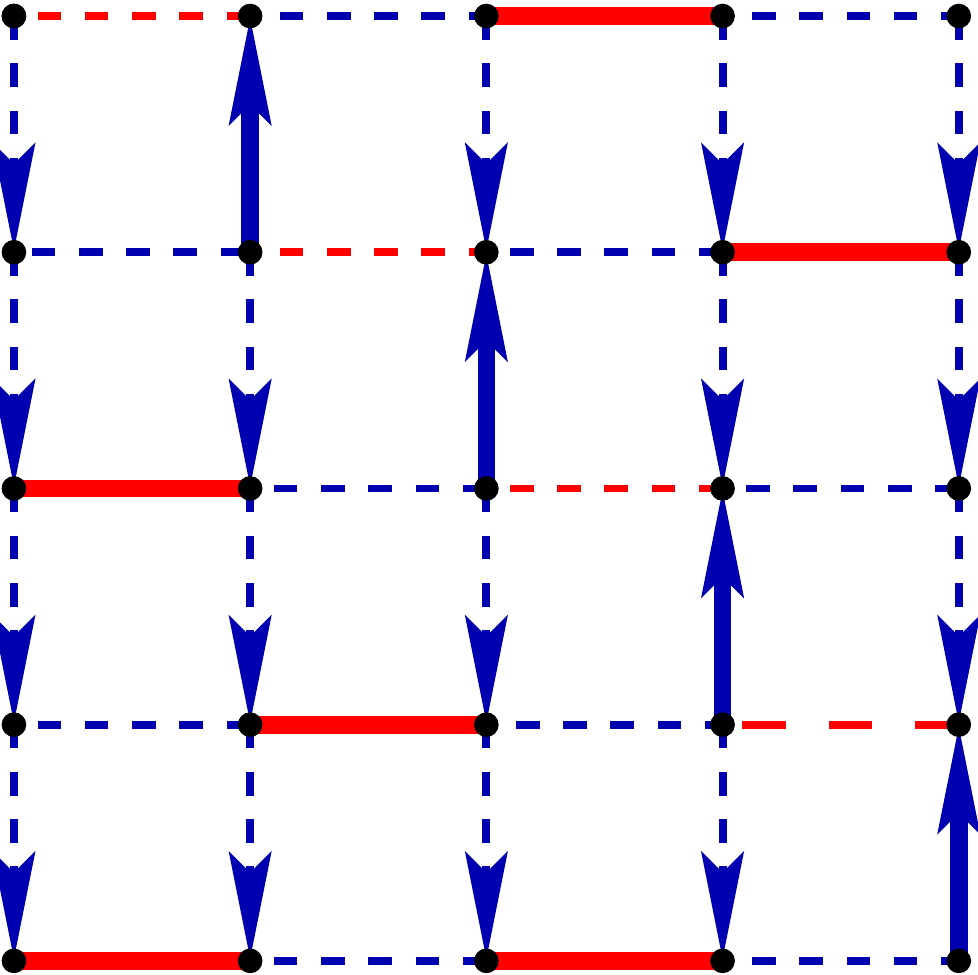}  
\hspace{1.0cm}
\includegraphics[width=0.26\textwidth]{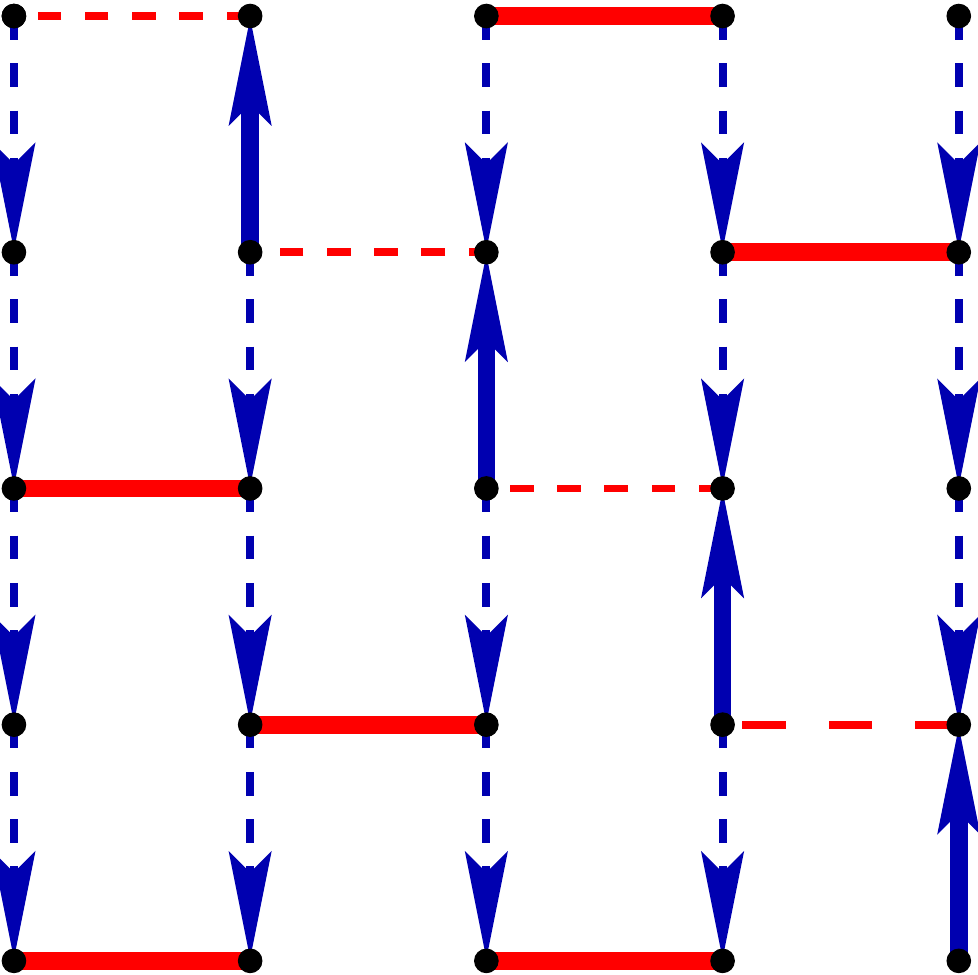}  
\caption{Schematic illustration of two anisotropic diamond-decorated square and honeycomb lattices (top panels) and their mapping to effective monomer-dimer models (bottom panels). The columnar and staggered configurations of the non-uniform bonds are shown in the left and middle panels. The right panels correspond to the brick-wall lattice that is topologically equivalent to the honeycomb lattice. The notation for the coupling constants on the diamond-decorated square lattices:
$J_1=1$ -- black bonds;
 $J_{2}^{H_2}$ -- red bonds from horizontal diamond units;
 $J_2^{V}$ and $J_2^{H_1}$ -- blue bonds from vertical and horizontal diamond units. 
For the honeycomb lattice, $J_2$ (red bonds) and $J'_2$ (blue bonds) are denoted for the diagonal couplings across the rhombi. Yellow shaded regions show elementary excitations for the three specific geometries.}
\label{fig:modela}
\end{figure*}

We complement the considerations of the main manuscript by investigating the square-lattice version of the model.
Like the diamond-decorated honeycomb lattice, also the square-lattice version exhibits a DT phase with
a macroscopically degenerate ground state \cite{Morita2016}. We first need to select a unique ground state among
these by favoring triplet formation on certain $J_2'$ dimer bonds.
Just making interactions spatially anisotropic \cite{Otsuka2009} does not suffice to completely lift the degeneracy in the case of the square lattice. Nevertheless,
the columnar or staggered $J_2'$ dimer patterns constitute two regular possibilities to favor a unique
ground state, see
the left and middle column of Fig.~\ref{fig:modela}, respectively. We will show below that the staggered dimer pattern leads to physics qualitatively similar to that of the diamond-decorated honeycomb lattice, whereas the columnar case does not exhibit signatures of Kasteleyn physics.

Throughout this Supplemental Material, we focus on the regime of strong dimer $J_2$ couplings ({\it i.e.}, $J_2^{V}, J_2^{H_1}, J_2^{H_2} > J_1$), for which the chosen parameters keep the system in the dimer–tetramer phase at zero temperature. Generalizing 
the procedure of our previous publication 
\cite{Karlova2024}, we obtain the energy of the monomer-dimer phase as follows
\begin{eqnarray}
	E^{(0)}_{\mathrm{MD}} = N_c\left(\varepsilon_{sd}^{V} +\frac{1}{2} \left(\varepsilon_{sd}^{H_1} +  \varepsilon_{sd}^{H_2} \right) \right)
	 = -\frac{3N_c}{8}\! \left(2 J_2^{V} + J_2^{H_1} + J_2^{H_2} \right),
	\label{eMD0}
\end{eqnarray} 
where $\varepsilon_{sd}^{V,H_1,H_2}=-\frac{3}{4}J_2^{V,H_1,H_2}$ denote the singlet energies on the corresponding dimers, 
and  $N_c=L_x L_y$ is the number of unit cells ($L_x$ and $L_y$ are the linear sizes of the square lattice in the $x$ and $y$ directions).
The energy of a singlet tetramer formed on a separate vertical, horizontal-1, or horizontal-2 diamond unit is likewise given by $\varepsilon_{st}^{V,H_1,H_2}=-2J_1+\frac{1}{4}J_2^{V,H_1,H_2}$. This allows us to define the excitation energies associated with promoting a singlet dimer to a singlet tetramer on each bond as follows
\begin{eqnarray}
	\Delta_{V,H_1,H_2} = \varepsilon_{st}^{V,H_1,H_2} - \varepsilon_{sd}^{V,H_1,H_2} = J_2^{V,H_1,H_2} - 2J_1.
\label{eq:excit}
\end{eqnarray}

We are interested in an effective low-temperature model restricted to singlet states  residing either on dimers or tetramers, which maps onto a monomer-dimer model on the square lattice. Within this mapping, a dimer between two nodal sites is assigned to a singlet-tetramer state on the corresponding diamond unit with one of the activities $x_1=e^{-\beta\Delta_{H_1}}$, $x_2=e^{-\beta\Delta_{H_2}}$, $y=e^{-\beta\Delta_V}$ depending on its orientation. 
Spins from nodal sites not covered by dimers contribute to the partition function through 
the degeneracy factor $z=2$.
The partition function of this effective model then takes the form
\begin{eqnarray}
	Z = e^{-\beta E^{(0)}_{\rm MD}}\sum_{\cal{C}} x_1^{H_1} x_2^{H_2} y^V z^M,
	\label{eq:Z_DM}
\end{eqnarray}
where $\sum_{\cal{C}}$ runs over all allowed monomer–dimer configurations on the square lattice and $M$, $V$, $H_1$, and $H_2$ denote the numbers of monomers, vertical dimers and two types of horizontal dimers constrained by $2(H_1+H_2+V)+M = N_c$. 

It is useful to introduce the rescaled parameters $\alpha_i=x_i/y=e^{-\beta(\Delta_{H_i}-\Delta_V)}$, 
$\gamma=z/y^{1/2}=2e^{\beta\Delta_V/2}$, 
which allow the partition function to be rewritten in the simplified form
\begin{eqnarray}
	Z = e^{-\beta E^{(0)}_{\rm MD}}y^{N_c/2} \sum_{\cal{C}} \alpha_1^{H_1} \alpha_2^{H_2} \gamma^M.
	\label{eq:Z_DM2}
\end{eqnarray}
The subsequent analysis follows the transfer-matrix formalism when introducing a transfer matrix $V_{i,i+1}$ connecting two neighboring rows $i$, $i+1$ of vertical bonds on the square lattice \cite{Lieb1967}. Vertical dimers on the square lattice are associated with spin-up states, while the absence of a dimer corresponds to a spin-down state. In this representation, the partition function \eqref{eq:Z_DM2} can be expressed as $Z=e^{-\beta E^{(0)}_{\rm MD}}y^{N_c/2}{\rm Tr} \left[ \prod_{i=1}^{L_y} V_{i,i+1}\right]$. Until this point, the procedure for both non-uniform models with  columnar and staggered anisotropies are identical. Differences arise only when specifying the explicit form of the transfer matrix for each case. 

\section{Columnar monomer-dimer model on the square lattice}
\label{ssec:col}

In the specific case of the columnar monomer-dimer model on the square lattice, the transfer matrix is independent of the row index and it can be expressed in operator form as a product of three factors ${\cal V}=V_3 V_2 V_1$
\begin{eqnarray}
	\label{eq:V_i}
	V_1 & = & \prod_{i=1}^{L_x} \sigma^x_i, \quad
	V_2  =  \exp\left(\gamma\sum_{i=1}^{L_x} \sigma^{-}_i\right), 	\quad
	V_3  =  \exp\left(\sum_{i=1}^{L_x/2} \left[\alpha_1\sigma^{-}_{2i-1}\sigma^{-}_{2i} + \alpha_2\sigma^{-}_{2i}\sigma^{-}_{2i+1}\right]\right),
\end{eqnarray}
where $\sigma^x_i$ and $\sigma^{\pm}_i$ denote the Pauli matrices. The operator $V_1$ encodes the constraint of close-packed vertical dimers, whereas the operator $V_2$ takes into account the contribution of monomers. The non-uniform $V_3$ incorporates  the alternating character of activities of the horizontal dimers. Since the transfer matrix is identical for any pair of neighboring rows, the partition function can be further simplified as $Z=e^{-\beta E^{(0)}_{\rm MD}}y^{N_c/2}{\rm Tr} {\cal V}^{L_y}$. In the limit $L_y\to\infty$, the free energy per unit cell takes the form
\begin{eqnarray}
	f &=& -\frac{1}{N\beta}\ln Z = 	E^{(0)}_{\rm MD} - \frac{\Delta_V}{2} - \frac{1}{\beta L_x}\ln\Lambda_{\max},
\label{eq:free_en}	
\end{eqnarray}
where $\Lambda_{\max}$ is the largest eigenvalue of ${\cal V}$.

\section{Staggered monomer-dimer model on the square lattice}
\label{ssec:stag}

The procedure becomes more involved for the staggered monomer-dimer model on the square lattice. Since the transfer matrix now depends explicitly on the row index, we introduce the notation $V_{i,i+1} = V^{\rm odd}$ if $i$ is an odd number and $V_{i,i+1} = V^{\rm even}$ if it is an even number. Consequently, 
$V^{{\rm odd}/{\rm even}} = V_3^{{\rm odd}/{\rm even}} V_2 V_1$, where $V_1$, $V_2$ are given by Eqs.~\eqref{eq:V_i} above and the operators $V_3^{{\rm odd}/{\rm even}}$ are defined as
\begin{eqnarray}
	\label{eq:V3_oe}
	V_3^{\rm odd}  =  \exp\left(\sum_{i=1}^{L_x/2} \left[\alpha_1\sigma^{-}_{2i-1}\sigma^{-}_{2i} + \alpha_2\sigma^{-}_{2i}\sigma^{-}_{2i+1}\right]\right), \quad
	V_3^{\rm even} & = & \exp\left(\sum_{i=1}^{L_x/2} \left[\alpha_2\sigma^{-}_{2i-1}\sigma^{-}_{2i} + \alpha_1\sigma^{-}_{2i}\sigma^{-}_{2i+1}\right]\right).
\end{eqnarray}
The total transfer matrix can be accordingly written as ${\cal V}_{total} = (V^{\rm even} V^{\rm odd})^{L_y/2} = {\cal V}^{L_y}$, where 
\begin{eqnarray}
	\label{eq:tm_stag}
	{\cal V}^2 = V_3^{\rm even} V_2 [V_3^{\rm odd}]^{\dagger} [V_2]^{\dagger}.
\end{eqnarray}
The free energy is then obtained from Eq.~\eqref{eq:free_en} by substituting therein the largest eigenvalue of the transfer matrix (\ref{eq:tm_stag}).

\section{Columnar and staggered dimer models on the square lattice: exact results}
\label{ssec:col}

In case of closely-packed dimers, the problem reduces to a pure dimer model without monomer contributions, which admits exact solutions (see, e.g., Refs.~\cite{Wu2006,Izmail2015,Chen2019}). As shown in Ref.~\cite{Izmail2015}, the columnar dimer model (corresponding to the checkerboard A-type models in their classification) exhibits a singularity only when all activities are equal, {\it i.e.}, $\Delta_V=\Delta_{H_1}=\Delta_{H_2}$. On the contrary, the staggered dimer model becomes non-analytic under the condition (see Eq.~(21) in Ref.~\cite{Wu2006}) 
\begin{eqnarray}
	e^{-\beta\Delta_{H_2}} = 2e^{-\beta\Delta_{V}} + e^{-\beta\Delta_{H_1}},
\end{eqnarray}
which leads to the following equation for the Kasteleyn transition temperature for $\Delta_V=\Delta_{H_1}$
\begin{eqnarray}
	T_K = \frac{\Delta_V-\Delta_{H_2}}{\ln 3}.
\end{eqnarray}

\section{Specific heat of columnar and staggered diamond-decorated square lattices}
\label{ssec:specheat}

\begin{figure}[t!]
    \centering
    \includegraphics[width=0.34\textwidth]{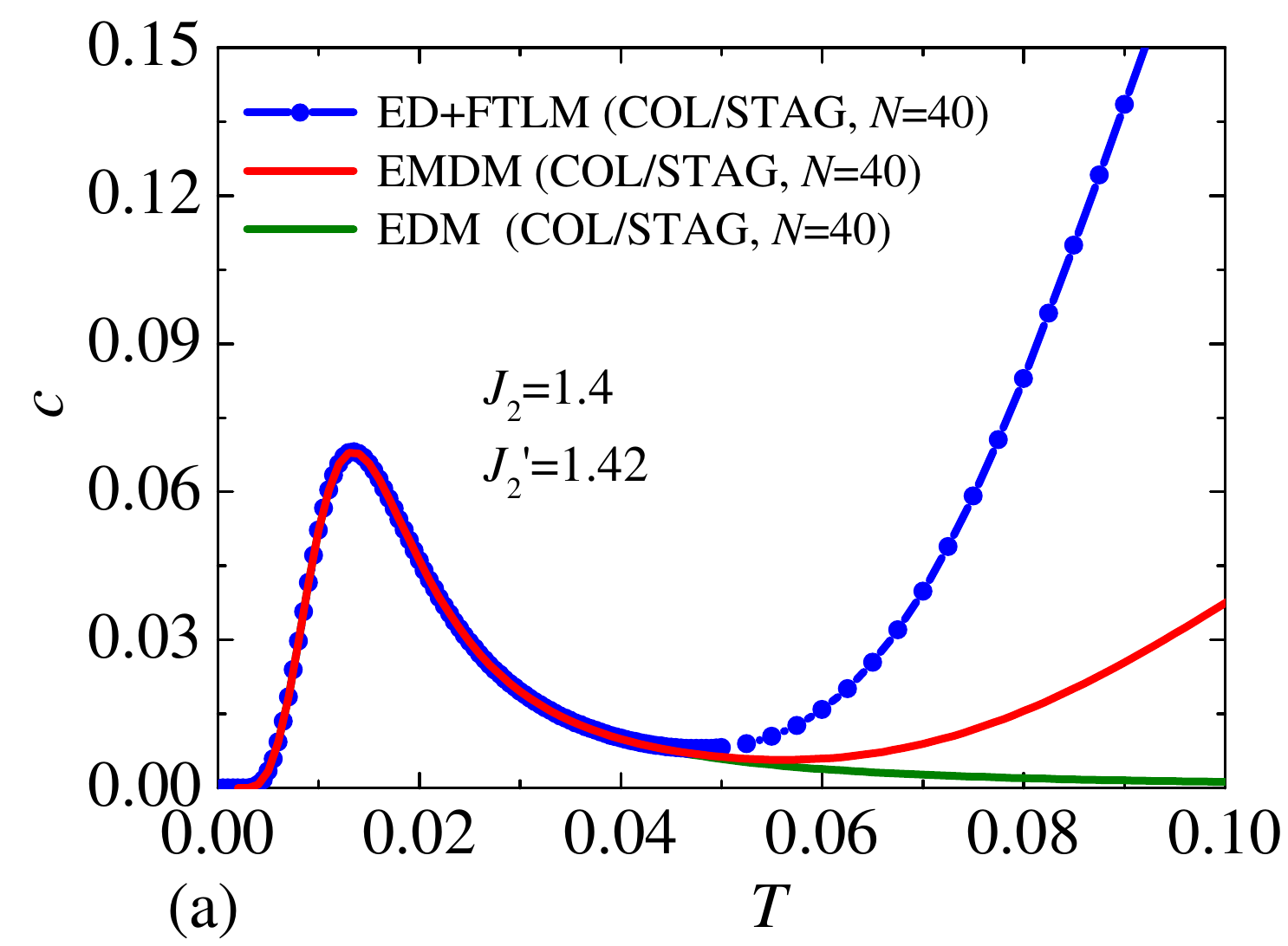}\hspace{-0.4cm}
		\includegraphics[width=0.34\textwidth]{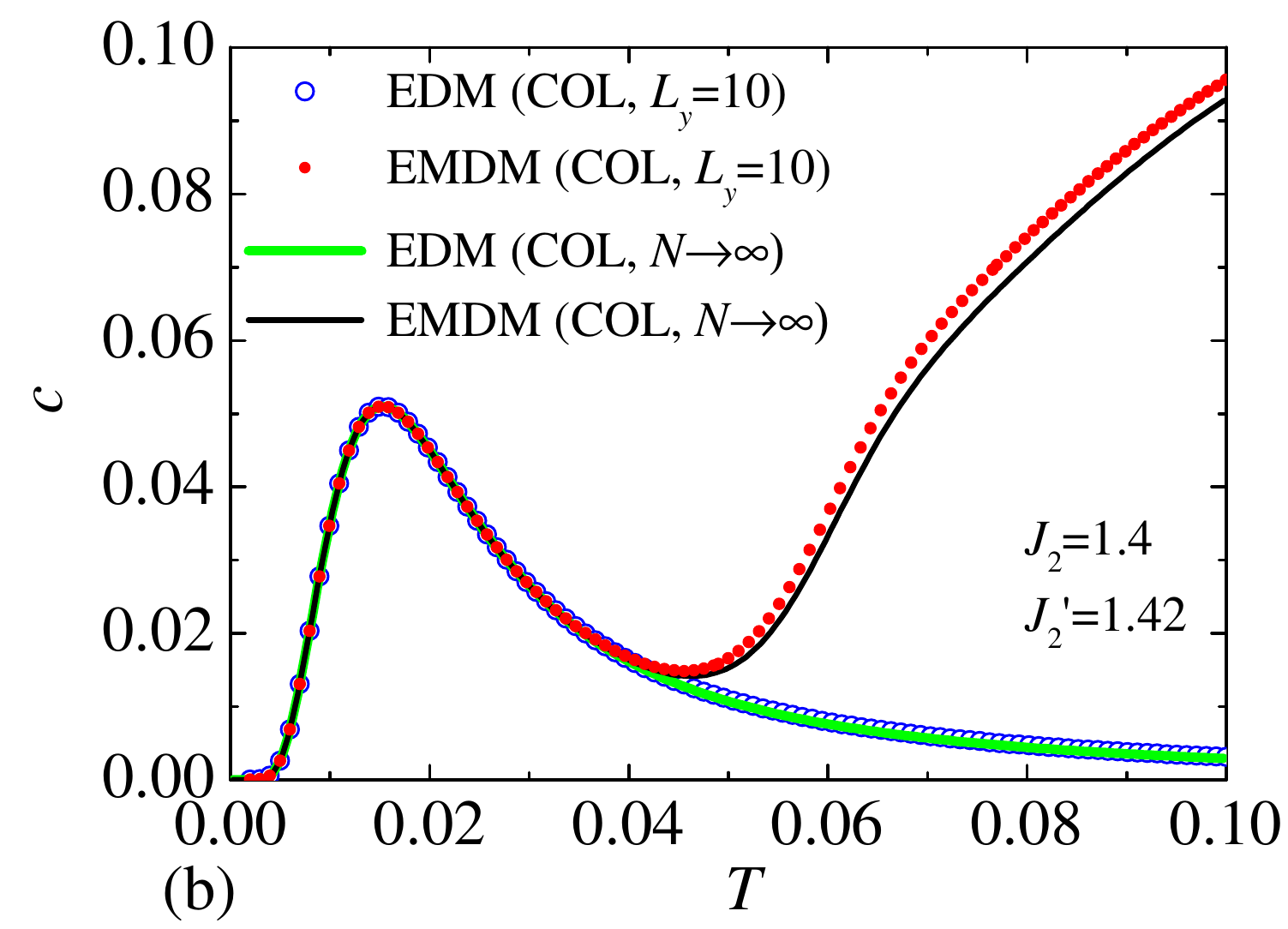}\hspace{-0.4cm}
		\includegraphics[width=0.34\textwidth]{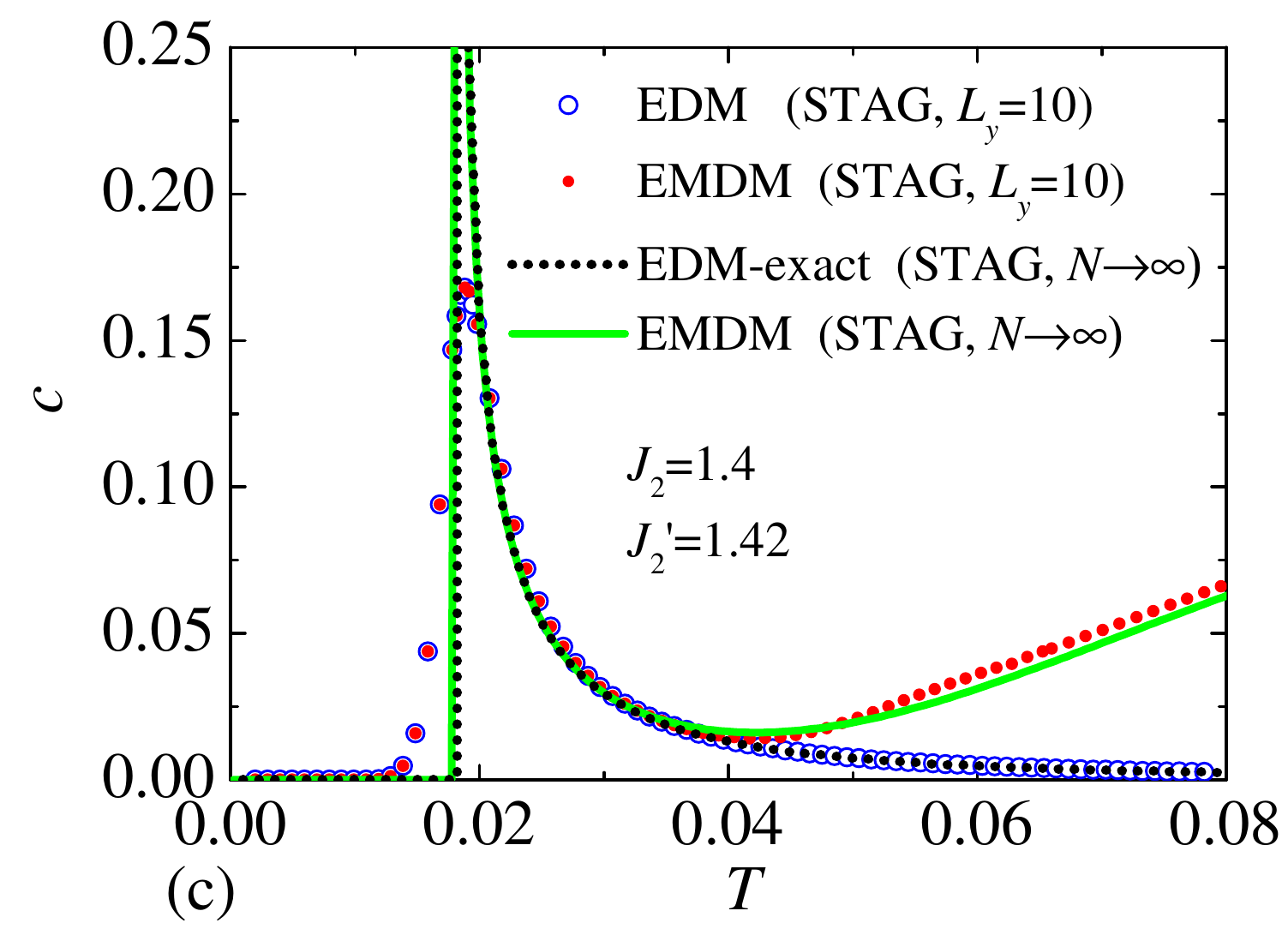}
    \caption{Specific heat per spin $c$ as a function of temperature $T$ for the anisotropic diamond-decorated square lattice with $J_2=1.4$ and $J_2'=1.42$. (a) Results for a system of $N=40$ spins, where the columnar (COL) and staggered (STAG) geometries are equivalent. ED+FTLM data for the original Heisenberg model are compared with the effective dimer model (EDM) and effective monomer–dimer model (EMDM), both obtained by exact numerical enumeration. The excellent agreement at low temperatures confirms the accuracy of the effective description. (b) Columnar  geometry: comparison between finite-width data ($L_y=10$) and the thermodynamic limit extracted from transfer-matrix (symbols) and CTMRG (solid lines) calculations. (c) Staggered geometry: same analysis as in (b), including the exact thermodynamic-limit result for the EDM (dotted line).}
    \label{fig:appendix_heat}
\end{figure}

To directly connect the above analytical considerations with numerical results, we now present a systematic comparison of the specific heat of the diamond-decorated square lattice with either columnar (COL) or staggered (STAG) arrangement of the diamond units.
Figure~\ref{fig:appendix_heat} summarizes the low-temperature behavior obtained from several complementary approaches: ED+FTLM data for the original Heisenberg model, exact enumeration within the effective dimer model (EDM) and the effective monomer–dimer model (EMDM) for finite clusters, transfer-matrix calculations for finite-width strips, and CTMRG results in the thermodynamic limit. This combined analysis allows us to distinguish the purely dimer-driven low-temperature anomalies characteristic of the columnar anisotropy from the Kasteleyn-type critical enhancement induced by the staggered anisotropy.

For a small system of $N=40$ spins, the columnar (COL) and staggered (STAG) anisotropic geometries become topologically equivalent~\footnote{Indeed, \emph{all} DT ground states are topologically equivalent
for the $N=40$ square lattice.
}. Consequently, both cases yield an identical 
heat capacity in this case. Figure~\ref{fig:appendix_heat}(a) shows that the specific heat exhibits a single pronounced low-temperature maximum. Notably, the ED+FTLM results for the original Heisenberg model agree exceptionally well with both the EDM and EMDM predictions in the vicinity of this peak. This agreement demonstrates that the low-temperature anomaly is dominated almost entirely by the dimer contributions. The monomer contributions become relevant only at temperatures $T\gtrsim 0.05$, which is approximately four times higher than the temperature at which the low-temperature peak is located. Hence, one may conclude that the observed low-temperature peak indeed originates from the dimer contributions.

We next analyze larger systems for the columnar geometry as shown in Fig.~\ref{fig:appendix_heat}(b). With increasing system size, the position of the low-temperature peak remains essentially unchanged.
However, the height of the peak is systematically reduced: while for the smaller system the peak approximately reaches the value of $c_{\rm max}\approx 0.07$, it decreases to about $c_{\rm max}\approx 0.05$ as the system size approaches the thermodynamic limit. This behavior indicates that, although the energy scale associated with the low-temperature anomaly is robust, its thermodynamic weight is progressively reduced with increasing system size. Furthermore, the EMDM starts to noticeably deviate from the EDM only at temperatures of the order of $T\approx 0.05$ well above the expected Kasteleyn temperature. This confirms that, even for larger systems, the low-temperature peak in the columnar case is entirely dominated by dimer excitations with monomer contributions remaining negligible in this temperature regime.
We also note that the peak in Fig.~\ref{fig:appendix_heat}(b) converges to a rounded form and does not
develop a singularity for $N \to \infty$.
Consequently, the columnar distortion does not give rise to a Kasteleyn-type critical enhancement of the specific heat.

A qualitatively different behavior emerges for the staggered geometry shown in Fig.~\ref{fig:appendix_heat}(c). As the system size increases, the position of the low-temperature peak again remains nearly unchanged, but in sharp contrast to the columnar case, it becomes sharper and its 
 height increases significantly. For instance, the transfer-matrix results on a $L_y=10$ strip show that the peak height increases from approximately $c_{\rm max}\approx 0.07$ to about $c_{\rm max}\approx 0.17$, {\it i.e.}, by nearly a factor of two. This trend is independently confirmed by the CTMRG data, which are in the thermodynamic limit essentially indistinguishable from the exact solution of the pure hard-dimer model on the square lattice up to temperatures $T\approx 0.04$, including a pronounced peak 
around the Kasteleyn temperature $T\approx 0.0182$. This quantitative agreement demonstrates that the staggered geometry exhibits the same low-temperature behavior as the anisotropic honeycomb lattice studied in the main text. The increasing low-temperature peak thus provides clear and unambiguous evidence for a Kasteleyn-type crossover in the staggered diamond-decorated square lattice originating from the proliferation of infinite string excitations. This mechanism is fundamentally distinct from the local excitations allowed in the columnar diamond-decorated case, c.f.\ yellow shaded regions in Fig.~\ref{fig:modela}(a) and (b).

\section{Mapping to the monomer-dimer model on the anisotropic honeycomb lattice}
\label{ssec:honeycomb}

The general procedure described above can be easily extended to the spin-1/2 Heisenberg model on the diamond-decorated honeycomb lattice studied in the main text. The effective monomer-dimer model for this geometry can be deduced directly from Eq.~\eqref{eq:Z_DM}, which provides the partition function for the staggered monomer-dimer model on the square lattice, by taking the limit $\Delta_{H_1}\to\infty$. This choice effectively removes the corresponding horizontal bond from the square lattice and reduces the square lattice to a brick-wall lattice that is topologically equivalent to the honeycomb lattice \cite{Grande2011}. 

The transfer matrix in the operator representation retains the form of Eq.~\eqref{eq:tm_stag} except that the operators 
$V_3^{{\rm odd}/{\rm even}}$ should be modified as follows
\begin{eqnarray}
	\label{eq:V3_honeycomb}
	V_3^{\rm odd}  =  \exp\left(\alpha_2 \sum_{i=1}^{L_x/2} \sigma^{-}_{2i}\sigma^{-}_{2i+1} \right), \quad \quad
	V_3^{\rm even}  =  \exp\left(\alpha_2 \sum_{i=1}^{L_x/2} \sigma^{-}_{2i-1}\sigma^{-}_{2i} \right).
\end{eqnarray}
To make the correspondence between the the spin-1/2 Heisenberg model on the diamond-decorated honeycomb lattice defined in the main text and Eq.~\eqref{eq:V3_honeycomb} explicit, we need to adjust the notation $\Delta_V = J'_2 - 2J_1$, $\Delta_{H_2} = J_2 - 2J_1$, and, consequently, $\alpha_2 =  e^{-\beta(J_2 - J'_2)}$.

\section{Anisotropic version of a fully frustrated kagome bilayer}
\label{ffkb}

\begin{figure*}
\centering
\includegraphics[width=\textwidth]{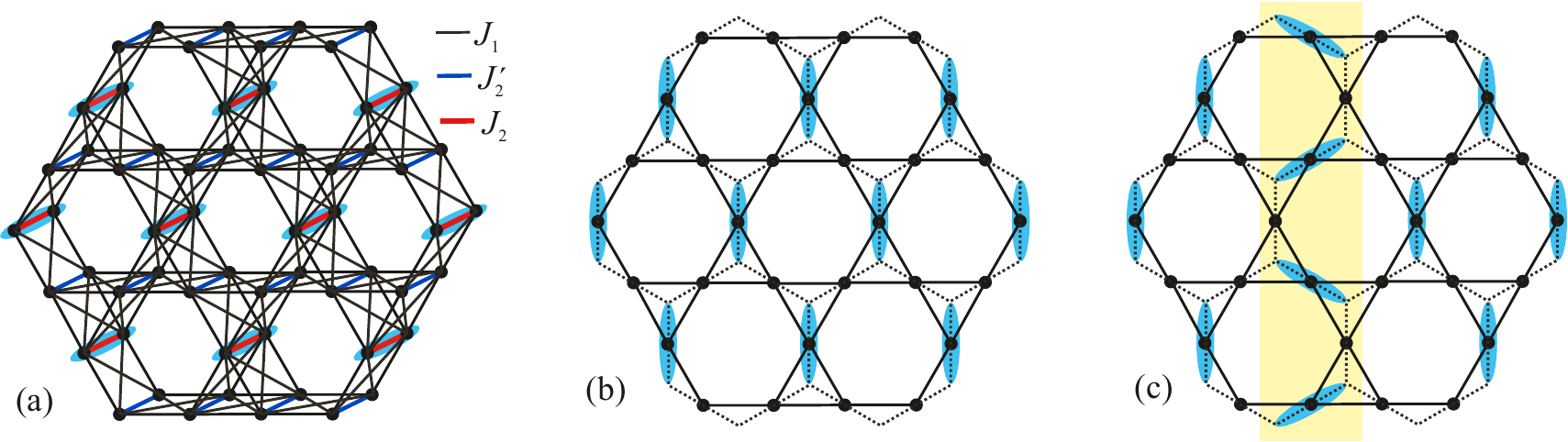}
\caption{(a) Anisotropic version of a fully frustrated kagome bilayer with three different coupling constants $J_1$, $J_2'$, and $J_2$ indicated by thin black, thick blue and red lines, respectively. Light-blue ovals denote dimer singlets forming the unique valence-bond-solid ground state, which is realized in the highly anisotropic regime $J_2 \gtrsim J_2' \gg J_1$. (b) Top view of the fully frustrated kagome bilayer and the corresponding mapping of the valence-bond-solid ground state onto a close-packed hard-dimer covering on the honeycomb lattice. (c) The highlighted strip illustrates an elementary string-like excitation generated from this exact ground state by successive rearrangements of the dimer singlets along the highlighted path.}
\label{fig:ffkb}
\end{figure*}

The mechanism leading to emergent Kasteleyn physics is not restricted to the distorted diamond-decorated honeycomb lattice considered in the main text and the staggered diamond-decorated square lattice  discussed in the Supplemental material. As a further illustrative example, we consider the anisotropic generalization of the fully frustrated kagome bilayer shown in Fig.~\ref{fig:ffkb}, whose isotropic counterpart $J_2=J_2'$ was recently investigated in Ref.~\cite{Yaremchuk2025}. In the highly frustrated regime $J_2 \gg J_1$, the isotropic fully frustrated kagome bilayer exhibits exact valence-bond-solid ground states composed of localized dimer-singlet states, which manifest themselves in the zero-temperature magnetization curve as zero, one-third, and two-thirds plateaus (see Fig.~2 of Ref.~\cite{Yaremchuk2025}). When a magnetic field stabilizes the intermediate two-thirds magnetization plateau, one third of the vertical dimers remains in a singlet state, whereas the remaining two thirds are fully polarized triplets. This gives rise to a macroscopically degenerate manifold of exact valence-bond-solid eigenstates subject only to the local constraint that each triangle of the kagome bilayer hosts a singlet-triplet-triplet configuration. This manifold of exact ground states can be mapped one-to-one onto close-packed dimers on the honeycomb lattice, where each singlet dimer corresponds to an occupied bond (dimer) of the effective honeycomb lattice shown in Fig.~\ref{fig:ffkb}(b) and (c) by dotted lines.

Introducing a weak spatial anisotropy $J_2 \gtrsim J_2' \gg J_1$ lifts the macroscopic degeneracy of these exact ground states in complete analogy with the distorted diamond-decorated honeycomb lattice and the staggered diamond-decorated square lattice. Under these conditions, the valence-bond-solid configuration schematically shown in Fig.~\ref{fig:ffkb}(a) becomes the unique ground state of the fully frustrated kagome bilayer, whose effective description in terms of a hard-dimer model on the honeycomb lattice is illustrated in Fig.~\ref{fig:ffkb}(b). In addition, the lowest-energy excitations above this ground state correspond to extended string defects as illustrated in Fig.~\ref{fig:ffkb}(c). These strings are generated by successive local rearrangements of neighboring singlet dimers and constitute the direct analogue of the string excitations responsible for the Kasteleyn transition in classical dimer models. Altogether, all essential features such as a macroscopically degenerate dimer manifold, anisotropy-induced selection of a unique ground state, and string-like excitations above it, are recovered in a frustrated quantum magnet with a completely different microscopic lattice geometry. 

Although a detailed thermodynamic analysis of this model is beyond the scope of the present work, the existence of the same constrained dimer manifold together with identical string-like excitations strongly suggests that it belongs to the same class of systems exhibiting emergent Kasteleyn criticality. This example demonstrates that the mechanism discussed in the present Letter is not restricted to a particular geometry of the diamond-decorated lattice, but it is a generic consequence of frustrated quantum magnets possessing an emergent dimer manifold with string-like excitations. These findings thereby substantially broaden the class of 2d frustrated quantum spin systems in which emergent Kasteleyn criticality may arise.

\bibliography{monomer-dimer}